\PassOptionsToPackage{table}{xcolor}
\documentclass[format=acmlarge, review=false, nonacm, screen=true, bookmarksnumbered, unicode, bookmarks=false]{acmart}

\usepackage[table]{xcolor} 
\usepackage[many]{tcolorbox}

\newcounter{example}
\definecolor{exHead}{RGB}{140,140,140}
\definecolor{exMain}{RGB}{220,220,220}

\newenvironment{example}[1][]{\refstepcounter{example}
\begin{tcolorbox}[enhanced,  colback=exMain, colframe=exHead, title= \textsf{\small{\textbf{Example~\theexample. #1}}}]
\setlength{\parindent}{1.5em}
   \rmfamily
   \footnotesize
   \par\medskip}
{  \medskip\end{tcolorbox}}
\usepackage{caption}
\usepackage{pifont}
\usepackage{booktabs}
\usepackage{multirow}
\usepackage{subcaption}
\usepackage[export]{adjustbox}

\usepackage{array, longtable}
\newcolumntype{L}[1]{>{\raggedright\let\newline\\\arraybackslash\hspace{0pt}}m{#1}}
\newcolumntype{C}[1]{>{\centering\let\newline\\\arraybackslash\hspace{0pt}}m{#1}}
\newcolumntype{R}[1]{>{\raggedleft\let\newline\\\arraybackslash\hspace{0pt}}m{#1}}

\newcommand{\smltick}{\footnotesize{\checkmark}}

\begin{document}

\title{The BIG Argument for AI Safety Cases}


\author{Ibrahim Habli}
\affiliation{%
  \institution{Centre for Assuring Autonomy, University of York}
  \city{York}
  \country{United Kingdom}}
\email{Ibrahim.Habli@york.ac.uk}

\author{Richard Hawkins}
\affiliation{%
  \institution{Centre for Assuring Autonomy, University of York}
  \city{York}
  \country{United Kingdom}}

\author{Colin Paterson}
\affiliation{%
  \institution{Centre for Assuring Autonomy, University of York}
  \city{York}
  \country{United Kingdom}}

\author{Philippa Ryan}
\affiliation{%
  \institution{Centre for Assuring Autonomy, University of York}
  \city{York}
  \country{United Kingdom}}

  \author{Yan Jia}
\affiliation{%
  \institution{Centre for Assuring Autonomy, University of York}
  \city{York}
  \country{United Kingdom}}

    \author{Mark Sujan}
\affiliation{%
  \institution{Centre for Assuring Autonomy, University of York}
  \city{York}
  \country{United Kingdom}}

    \author{John McDermid}
\affiliation{%
  \institution{Centre for Assuring Autonomy, University of York}
  \city{York}
  \country{United Kingdom}}

\renewcommand{\shortauthors}{Habli et al.}

\begin{abstract}

\textbf{Abstract. }We present our Balanced, Integrated and Grounded (BIG) argument for assuring the safety of AI systems. The BIG argument adopts a whole-system approach to constructing a safety case for AI systems of varying capability, autonomy and criticality. Firstly, it is balanced by addressing safety alongside other critical ethical issues such as privacy and equity, acknowledging complexities and trade-offs in the broader societal impact of AI. Secondly, it is integrated by bringing together the social, ethical and technical aspects of safety assurance in a way that is traceable and accountable. Thirdly, it is grounded in long-established safety norms and practices, such as being sensitive to context and maintaining risk proportionality. Whether the AI capability is narrow and constrained or general-purpose and powered by a frontier or foundational model, the BIG argument insists on a systematic treatment of safety. Further, it places a particular focus on the novel hazardous behaviours emerging from the advanced capabilities of frontier AI models and the open contexts in which they are rapidly being deployed. These complex issues are considered within a wider AI safety case, approaching assurance from both technical and sociotechnical perspectives. Examples illustrating the use of the BIG argument are provided throughout the paper.
\end{abstract}


\maketitle

\section{Introduction}

AI is increasingly recognised for its potential to deliver significant benefits at the individual and societal levels, cutting across sectors and national boundaries \cite{UNAI}. AI’s use in Healthcare is a classic example \cite{topol2019high}. Alleviating pressures on healthcare systems across the world is a global priority. Latest OECD reports highlight AI’s capacity to “\textit{automate administrative tasks and free substantial time for healthcare providers to focus on patient care}” \cite{OECD1}. In particular, AI can extend\textit{ “health services to remote or underserved areas, improving healthcare access for millions worldwide living hours away from the nearest healthcare facility}”\cite{OECD1}.

However, there have been major concerns about the harms, e.g. physical and psychological \cite{zeng2024ai} \cite{lee2025saif}, that the use of AI could cause, especially when the technology is embedded into wider engineered systems and complex social settings \cite{habli2024ai}\cite{uuk2024taxonomysystemicrisksgeneralpurpose}\cite{chandra2024livedexperienceinsightunpacking}. For instance, accidents and incidents involving autonomous driving have been newsworthy \cite{stilgoe2021can}. Two notable examples are an Uber self-driving car crash that led to the death of Elaine Herzberg in Tempe, Arizona in 2018 \cite{uber}, and the suspending of Cruise ‘robotaxi’ operations, a subsidiary of General Motors, in San Francisco, California in 2023 following a series of pedestrian injuries \cite{koopman2024anatomy}.

Since the publication of Google’s landmark paper “\textit{Attention is all you need}” in 2017 \cite{vaswani2017attention} and the subsequent rise of Large Language Models (LLMs) \cite{steyvers2025large}, the scope in which AI is being deployed within safety-critical systems has grown significantly. The rapid scaling of capabilities, driven by increasing data and computational resources, shows no signs of slowing down \cite{wu2024inference}\cite{bengio2025international}. Further, the release of R1 in January 2025, an LLM developed by DeepSeek, suggests that computational techniques beyond scaling pre-training data could lead to more efficient results \cite{gibney2025china}. Interestingly, work is also well underway to incorporate LLMs as well as multimodal variants, such as vision language models, in self-driving vehicles \cite{xie2025vlms} \cite{yang2023llm4drive}. Vision Language Models (VLMs) use the same transformer-based approach as LLMs but with images rather than text as inputs. Indeed, for safety-critical applications, the tendency of LLMs to produce factually incorrect but seemingly plausible outputs, i.e. AI hallucinations  \cite{farquhar2024detecting}\cite{jones2025ai}, presents an open challenge for their use in such applications.

The scale and significance of AI-induced harms make proactive and systematic assurance of safety a priority \cite{fearnleyrisk}\cite{hendrycks2023overviewcatastrophicairisks}\cite{10.1145/3600211.3604673}. This is particularly the case as AI systems are granted greater autonomy \cite{burton2020mind}\cite{chan2023harms} and the use of powerful, general-purpose foundational models, or frontier AI, increases \cite{bengio2023managing} \cite{bengio2025international} \cite{lazar2023ai}. For instance, critical tasks such as clinical diagnoses \cite{habli2020artificial} or driving in open environments \cite{koopman2024redefining} are gradually being delegated to AI-based systems. Up until now, the partial automation of such activities has assumed, and is often built around, human oversight as a central risk mitigation measure \cite{porter2018moral}. The transition of such tasks from human to AI fundamentally challenges long-established engineering standards and safety practices, which centre on high degrees of control of systems and their environments \cite{bainbridge1983ironies}\cite{leveson2016engineering}\cite{monkhouse2020enhanced}.  It is worth noting that such challenges are also present with the widespread introduction of AI even when human oversight is assumed, due to both technical reasons (e.g. lack of transparency in AI decision making) as well as socio-technical reasons (e.g. an organisational culture where people lack psychological safety to disagree with AI decision making, which hinders their ability to exercise oversight) \cite{sujan2019human}.  

For six decades, safety cases have been an accepted means for assuring the development, deployment, maintenance and decommissioning of safety-critical systems across many sectors, most notably nuclear, defence and automotive \cite{kelly1999arguing}\cite{bishop2000methodology}\cite{sujan2016should}. Safety cases provide a way to communicate a clear, comprehensive and defensible argument that a system is acceptably safe to operate in a given context \cite{m2017a}. They consist of a structured argument, supported by a rigorous body of evidence.  Importantly, the use of safety cases represents a shared mindset and understanding of how safety should be managed and evidenced, extending beyond merely viewing the safety case as a document \cite{sujan2024changing}. There is growing interest in appraising the suitability of safety cases for supporting the assurance of AI-based systems and services \cite{hawkins2021guidance} \cite{buhl2024safety}. There has been substantial research undertaken into the development of compelling safety cases for narrow AI models, especially those used in highly autonomous applications \cite{burton2017making}. Further, with the remarkable advancement of frontier AI models, particularly LLMs, there has been growing interest in the development of safety cases for such general-purpose AI systems \cite{clymer2024safety}. This has focused almost exclusively on the consideration and mitigation of large-scale risks associated with the development of unauthorised and unacceptable capabilities of LLMs \cite{goemans2024safety}. This work is significant and essential. However, major gaps remain to integrate this fast emerging work into the wider consideration of immediate and imminent risks of harm to individuals and society caused by the integration of general-purpose AI models into safety-critical products and services.

In particular, as with any technology \cite{moller2018handbook}, it is essential to consider the risk of harm when using a general-purpose AI as part of a wider system to undertake specific tasks. The safety cases created must take into account the broad operational context to ensure that different kinds of hazards, both immediate and long-term, are addressed in an integrated and consistent manner. In this way, the safety cases are grounded in the established principles of system safety that are expected for all safety-related applications \cite{kelly2004systematic}.

In this paper, we propose a systematic approach to assuring the safety of AI through the provision of \textit{a whole system safety case.} Specifically, this paper proposes a \textit{Balanced, Integrated and Grounded (BIG) argument} that addresses AI safety at both the technical and the sociotechnical levels. Firstly, it balances safety with other critical ethical issues, such as privacy and equity, acknowledging the complexities and trade-offs in the broader societal impact of AI. Secondly, it integrates the social, ethical and technical aspects of safety assurance in a traceable and accountable manner. Thirdly, it is grounded in long-established norms and practices from safety-critical systems, such as being sensitive to context and maintaining risk proportionality. The approach takes an architectural approach to AI safety cases, demonstrating how an overall AI safety argument, in its different social, ethical and engineering aspects, comes together. 

The BIG argument primarily builds on the following patterns and methodologies:
\begin{itemize}
    \item Principles-based Ethics Assurance (PRAISE) \cite{porter2024principles}
    \item Assurance of Autonomous Systems in Complex Environments (SACE) \cite{hawkins2022guidance}
    \item Assurance of Machine Learning for use in Autonomous Systems (AMLAS) \cite{hawkins2021guidance}
    \item Emerging safety argument patterns for frontier AI models (e.g. \cite{goemans2024safety}\cite{korbak2025sketch})
\end{itemize}
Overall, the goal of the BIG argument is to improve transparency and accountability for the safety of AI, and ultimately contribute to the development, deployment and maintenance of justifiably safe AI systems. The target audience of this work is diverse:
\begin{itemize}
    \item Safety specialists and regulators, including those who have to sign-off safety cases;
    \item AI developers and system/software engineers, covering the full engineering lifecycle;
    \item Non-technical professionals, including managers, lawyers, ethicists, policy makers and social scientists; and
    \item Users, varying from trained professionals, e.g. clinicians and pilots, to the general public whose safety is at stake.
\end{itemize}

The paper is organised as follows. We first provide an overview of safety cases. Next, we present our BIG safety argument, followed by a detailed description of its core sub-arguments (ethical/social, engineering and technical). To illustrate its application, we include examples from diverse domains and AI technologies of different capability, autonomy and criticality. Finally, we conclude with overarching themes for future work.

\section{A Very Brief Overview of Safety Cases}
A highly-cited definition of safety cases comes from the UK Defence Standard 00-56, in which a safety case is described “\textit{a structured argument, supported by evidence, intended to justify that a system is acceptably safe for a specific application in a specific operating environment} \cite{m2017a}”. The more critical, novel and uncertain the system and its context, the more detailed the safety argument and evidence are expected to be \cite{habli2021safety}.

Safety cases were first adopted in the UK nuclear industry in 1965 in response to the Windscale fire accident in 1957 \cite{arnold2016windscale}. The adoption of safety cases was part of a shift in regulation from compliance-based to goal-based approaches. The thinking behind this is that while the regulator can set goals, the developers and operators of safety-critical systems are best placed to determine the means through which these regulatory goals can be achieved, especially in industries that are fast changing and where prescriptive standards would be at risk of lagging behind the state of the art \cite{bishop2004future}\cite{kelly2005goal}. Following major accidents in other industries, the goal-based regulatory approach supported by safety cases was adopted across UK safety-critical industries, including offshore oil and gas production (Piper Alpha explosion in 1988 \cite{CullenPiper}) and railways (King's Cross escalator fire in 1987 \cite{Kingscross} and Clapham main line derailment in 1988 \cite{Clapham}). The construction industry is the most recent domain (high-rise residential buildings) to adopt safety cases following the Grenfell Tower fire in 2017, where 72 people died \cite{Grenfell}.

When used as a proactive approach to safety management, the use of safety cases has the potential to offer significant benefits \cite{sujan2016should}. Most importantly, safety cases can provide assurance to developers and operators of safety-critical systems that they have properly understood relevant risks and that these are sufficiently managed \cite{CullenPiper}.  In addition, safety cases as an approach should not be mistaken as simply a written document, i.e. a safety case report, but rather as a structured way and shared understanding of thinking about and managing safety \cite{sujan2024changing}. The mindset underpinning the safety case approach is arguably its greatest strength as it encourages proactive, continual engagement with safety in an open and transparent manner.  

However, when safety cases are applied without the accompanying shift in mindset, there is an acknowledged risk that the approach can degenerate into a paper-based bureaucratic exercise that offers little towards improving safety \cite{kelly2008safety}. This was highlighted in the review following the loss of a Royal Air Force Nimrod aircraft in Afghanistan in 2006 \cite{haddon2009nimrod}. Furthermore, critics of the safety case approach have highlighted that there is a lack of robust evidence about the effectiveness of safety cases as a regulatory approach \cite{leveson2011use}\cite{graydon2020towards}\cite{rinehart2017understanding}.  In practice, safety cases are adopted based on face validity rather than conclusive evidence of their utility \cite{sujan2016should}.    

For over three decades, research in safety cases, mostly conducted in collaboration with industry, has focused on several key areas:
\begin{itemize}
    \item Notations: Mostly graphical representations (and also others \cite{holloway2008safety}\cite{holloway2023friendly}), primarily using the Goal Structuring Notation (GSN) \cite{assurance2018goal} and Claim-Argument-Evidence (CAE) \cite{netkachova2015tool};
    \item Processes: Integrating safety assurance into design from the earliest stages and throughout \cite{graydon2007assurance};
    \item Model-based engineering: Providing metamodels and ontologies for model-based safety cases, utilising model transformation, traceability and validation \cite{wei2019model};
    \item Safety case automation: Tool support for argument generation, integration and evaluation \cite{denney2018tool}\cite{carlan2022automating};
    \item Formal reasoning: Representing safety arguments in mathematical formats to improve precision and consistency \cite{rushby2009formalism};
    \item Modular safety cases: Promoting separation of concerns and enabling compositional representation and analysis \cite{kelly2001concepts}\cite{fenn2007safety};
    \item Confidence assessment: Evaluating different kinds of uncertainty, both qualitatively and quantitatively \cite{denney2011towards}\cite{hawkins2011new}\cite{goodenough2012toward}\cite{bloomfield2024confidence}\cite{guiochet2015model};
    \item Reuse: Promoted through argument patterns and templates \cite{kelly1997safety}\cite{ye2005justifying};
    \item Evolution and updates: Addressed via the concept of dynamic safety cases and phased safety case development \cite{denney2015dynamic}\cite{calinescu2017engineering}\cite{asaadi2020dynamic}\cite{codaf_safecomp}; 
    \item Argument-based assurance: Extended to cover properties beyond safety \cite{rushby2015interpretation}\cite{holloway2015explicate}, such as security \cite{mohamad2021security}\cite{alexander2011security}, ethics \cite{porter2024principles}, and trustworthiness \cite{burr2023ethical}.
\end{itemize}

The current literature already covers some methods and case studies on the use of safety cases for AI systems.  Most of these studies often have a narrow focus, such as a tightly-scoped technical problem (e.g. robustness of neural networks for pedestrian detection \cite{gauerhof2020assuring}) or a specific legal consideration (e.g. liability for misdiagnosis using a clinical AI tool \cite{lawton2024clinicians}\cite{birch2022clinical}). Having a narrow focus offers advantages, such as providing stronger and more detailed evaluation evidence. These studies are typically driven by specific requirements in safety standards or guidelines for producing safety cases, most notably in automotive \cite{burton2023addressing}\cite{koopman2023ul}\cite{borg2023ergo}, healthcare \cite{jia2021safety} and aviation \cite{denney2023assurance}\cite{fenn2023architecting}.

More recently, safety cases have been considered in the context of frontier AI assurance. Unlike existing literature on safety cases that looks at AI systems in specific contexts and for clearly defined purposes (i.e. downstream AI safety \cite{mcdermid2024upstream}), the emerging work on frontier AI safety cases takes a capability-based approach in isolation of a specific deployment environment (i.e. upstream AI safety) \cite{clymer2024safety}\cite{buhl2024safety}. 
The main questions of interest here are: Does a frontier AI model pose a dangerous threat or hazard to society, and is it controlled or controllable? A key focus of these arguments is capability misuse by ‘\textit{bad actors}’, say for the purpose of compromising cybersecurity or producing bioweapons leading to catastrophic or even existential harm. 

While the term "capability" is widely used in the literature, a concrete definition is rarely provided. Instead, works commonly present a limited set of examples of capabilities, leaving readers to infer their own definition. Without a clear definition, it becomes difficult to develop approaches to tackle this important problem or determine whether solutions would apply to capabilities more broadly.
One definition of capabilities has been supplied in the 2025 International AI Safety Report ~\cite{ISR-Safety} as ``\textit{The range of tasks or functions that an AI system can perform and the proficiency with which it can perform them}". 
This definition highlights the need for a system-centric approach to AI safety. AI systems are typically orchestrations of frontier models, machine learning components, traditional software, hardware and human operators. As such, assuring the safety of such systems requires more than a consideration of the technological features of AI models. We must also consider the context into which they are to be deployed, as well as the social and ethical landscape. What is needed is a balanced, integrated and grounded argument.

\section{The Balanced, Integrated and Grounded (BIG) Argument}

The fragmented state of the literature on AI safety does not help address a central question \cite{bengio2025international}: how does the whole safety argument come together?

\begin{figure}
  \centering
  \includegraphics[width=0.9\textwidth]{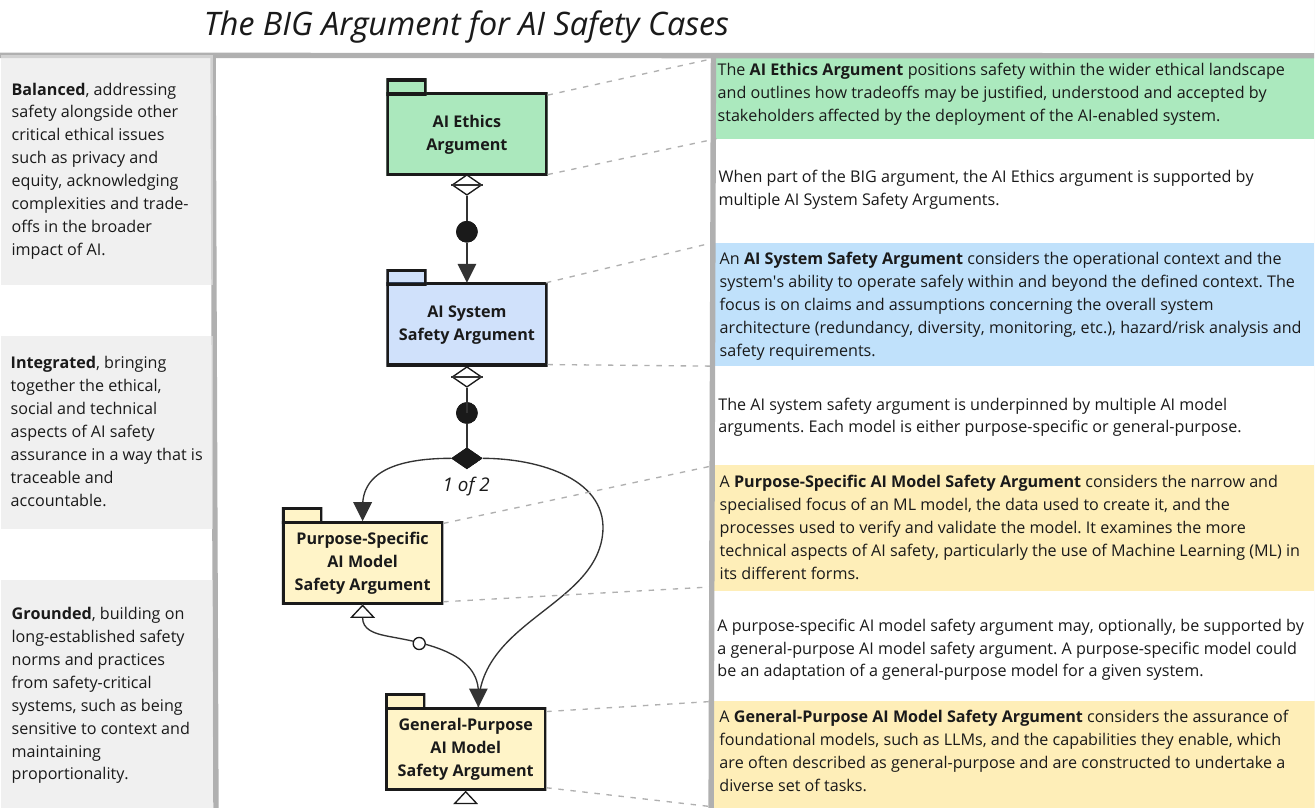}
  \caption{\label{fig:BIGArg}The Balanced, Integrated and Grounded (BIG) argument, represented in GSN}
\end{figure}

We address this challenge by proposing our Balanced, Integrated and Grounded (BIG) argument for AI safety cases. It is centred on three key characteristics:
\begin{enumerate}
    \item \textbf{B}\textbf{alanced argument:}  Protection from or avoidance of harm is a fundamental ethical principle enshrined in professional codes and legal standards. This principle of safety, or  \textit{non-maleficence}, is one part of a set of core principles that should guide critical decision-making about complex societal and technological interventions. Balancing these principles often requires careful consideration and trade-offs  \cite{hansson2018perform}\cite{porter2024principles}. For AI systems, a clear and explicit safety argument should facilitate a robust and inclusive dialogue among relevant stakeholders or their representatives, addressing their perspective on safety risks, relating them to other critical issues such as privacy and bias \cite{bender2021dangers}\cite{ganapathi2022tackling}. This dialogue should form the basis for a proportionate approach to risk reduction, ensuring that AI safety is not unjustifiably achieved at the expense of other ethical values like fairness \cite{leslie2024ai} and human autonomy \cite{prunkl2024human}.
    \item \textbf{I}\textbf{ntegrated argument:} The technical side of AI safety cannot be considered in isolation. Safety assurance must integrate the relevant ethical, social and engineering dimensions of AI deployment \cite{dobbe2022system}. Claims, assumptions and evidence for safety in one dimension can refine those in another. Consider a clinical reinforcement learning system used to support sepsis treatment in intensive care units \cite{jia2021safety}\cite{komorowski2018artificial}. Claims about mitigating the \textit{clinical} \textit{risk} of adverse outcomes, such as unnecessarily recommending a vasopressor for a patient with high blood pressure, are intertwined with the \textit{technical objective} of penalising the underlying model for exhibiting such risky behaviour. These clinical and technical issues are interconnected and should be reasoned about in a traceable and integrated manner.
    \item \textbf{G}\textbf{rounded argument:} Key aspects of AI assurance will focus on unique issues related to the AI lifecycle and design, such as the completeness of training and testing datasets, model performance and robustness and output explainability \cite{ashmore_assuring_2021}. However, to be meaningful in a safety context, these issues must be linked to and grounded in established safety norms \cite{rasmussen1997risk} such as hazard-oriented safety requirements, risk-driven controls \cite{leveson2023introduction} and just safety cultures \cite{dekker2016just}\cite{dekker2019foundations}. While AI presents significant and novel safety challenges, these traditional norms and practices remain relevant. In fact, the complex nature of AI systems further emphasises the need for rigour \cite{rae2020manifesto} and transparency \cite{habli2020enhancing} in how risk acceptability \cite{lowrance1976acceptable}\cite{hansson2003ethical} and safety requirements are argued about, reviewed and challenged \cite{sujan2017can}.
\end{enumerate}


\section{Sketching the BIG Argument}

Figure \ref{fig:BIGArg} sketches our BIG AI safety argument. The argument structure, represented using the patterns and modular features of the Goal Structuring Notation (GSN), incorporates the following sub-arguments:

\begin{itemize}
    \item \textbf{AI Ethics Argument:} This argument considers safety amongst other principles that must be assured for the ethical use of AI. Trade-offs between these principles will often be inevitable. The argument outlines how tradeoff decisions and assumptions may be justified, understood and accepted by affected stakeholders or their representatives (e.g. regulators). Our Principles-based Ethics assurance (PRAISE) patterns provide a basis for this argument \cite{porter2024principles}.
    \item \textbf{AI System Safety Argument:} This covers the wider system in which an AI model is deployed, which could be physical (e.g. an autonomous vehicle) or procedural (e.g. drug discovery). The argument details claims and assumptions about relevant hazards and risks, and evidence for the suitability of system-level risk controls including redundancy, diversity, monitoring and meaningful human control/oversight. Our Safety Assurance of Autonomous Systems in Complex Environments (SACE) patterns provide a basis for this argument \cite{hawkins2022guidance}.
    \item \textbf{Purpose-specific AI Model Safety Argument:} This represents claims and evidence for AI models trained and tested for a specific purpose, such as identifying pedestrians using a convolutional neural network or recommending a treatment using reinforcement learning. A key focus of the argument is the justification of the training and testing datasets and the allocated safety requirements and metrics. Our Assurance of Machine Learning for use in Autonomous Systems (AMLAS) patterns provide a basis for this argument \cite{hawkins2021guidance}.
    \item \textbf{General-Purpose AI Model\footnote{Here we use the term General-Purpose AI as defined in~\cite{GenAI-def} i.e. a  model created without an explicit consideration of the final system within which the model will be deployed. Such models can typically, without substantial modification, be \textit{tuned} to meet the needs of a specific role within a system.} Safety Argument: }This represents safety claims about general-purpose models and capability-specific risks and guardrails, and the supporting evidence, particularly from evaluations (evals), independent audits and red teaming \cite{wei2023jailbroken}\cite{perez2022red}. Preliminary argument templates and example safety cases have started to emerge, with focus on specific capabilities, e.g. the ability of a frontier AI to hide its behaviours until deployment or undermine oversight \cite{AnthropicSketches}\cite{balesni2024towards}. A recent report by the UK AI Security Institute stated that “\textit{Frontier AI safety cases should make arguments about more than just the technical system}” \cite{AISIUK2025}. The BIG argument advances this proposition. It offers traceable means for integrating the assurance of the technical capabilities of these advanced models with a wider set of sociotechnical issues at both the system and societal levels \cite{dobbe2024toward}\cite{lazar2023ai}\cite{weidinger2023sociotechnical}.
    
\end{itemize}

It is important to emphasise that when these General-Purpose AI capabilities are made explicit,\footnote{In~\cite{DeepMind-CCL-def}, a capability is suggested as ``\textit{Capable of significantly enabling a non-expert to develop known biothreats that could increase their ability to cause severe harm compared to other means}". Here the role of the model within the larger system may be considered as a  publicly accessible knowledge source, or natural language search tool. For harm, of this type, to occur requires a societal context in which the non-expert has an adversarial mindset and access to the resources necessary to make use of the knowledge supplied. To mitigate this issue at the technical level requires a consideration of issues such as data poisoning and deployment mitigation strategies in the wider software framework. The General-Purpose AI model does not, on its own, have the capability of producing harm.} they often conflate societal and systems thinking rather than merely considering the technical capabilities of the model. This conflation must be disentangled through a structured argument, as illustrated by the BIG argument, before sufficient action can be taken to address the arising safety issues.

\begin{figure}
  \centering
  \includegraphics[width=0.9\textwidth]{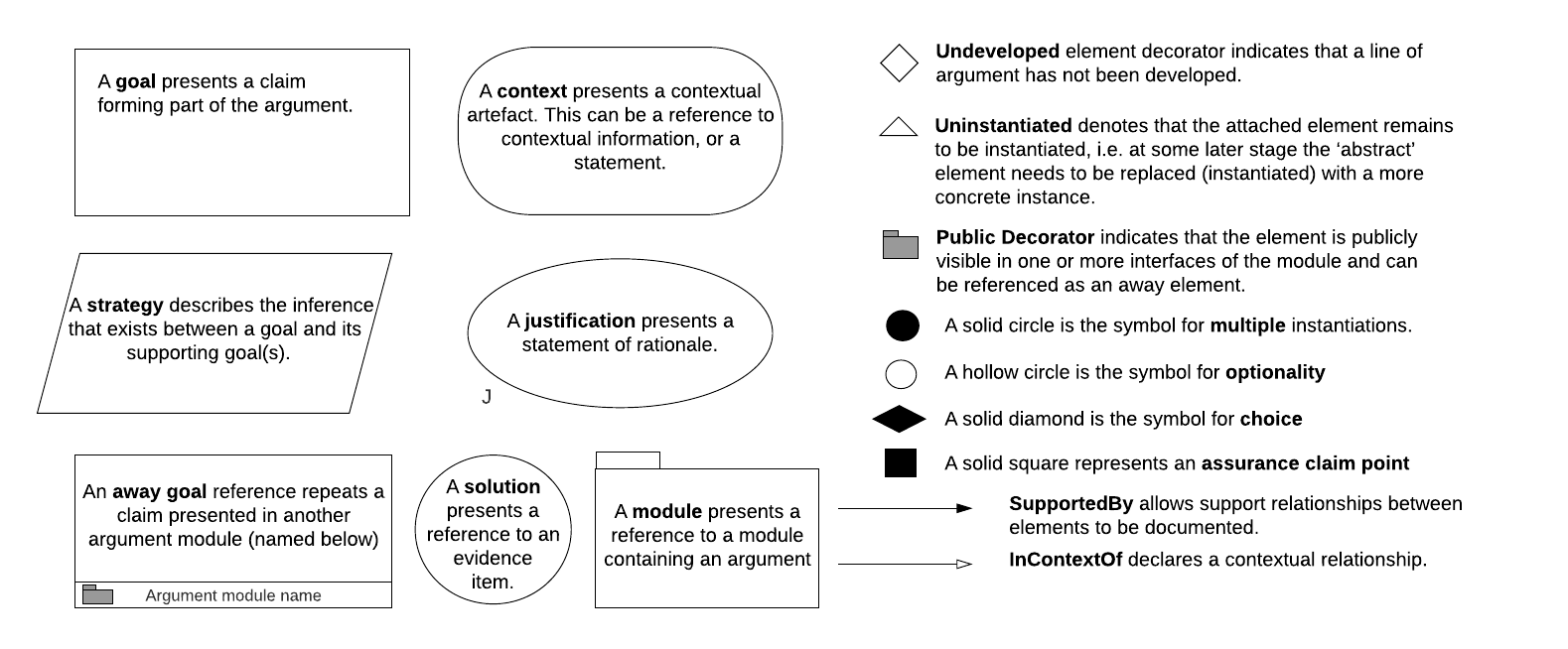}
  \caption{\label{fig:GSNLeng}Symbols and elements of a GSN argument. Extracted and adapted from Assurance Case Working Group \cite{assurance2018goal}}
\end{figure}

Structurally and notationally, the overall GSN argument pattern is described as follows (see Figure \ref{fig:GSNLeng} for a GSN legend): The \textit{AI Ethics Argument module} is supported by one or more \textit{AI System Safety Argument modules} (the multiplicity is represented by the solid circle). This indicates that multiple AI systems may require separate safety arguments within a broader safety case. For example, a ground-based system and an airborne system might each need a separate safety argument as part of an overall aviation safety case for an AI-enabled navigation capability. These modules are also ‘undeveloped’ and ‘uninstantiated’. This means that the argument contained within the modules requires further support and specific details (hence its status as a template). 

The multiplicity and choice symbols (solid diamonds) under the\textit{ AI System Safety Argument module} indicate that this module may be supported by one or more argument modules corresponding to either general-purpose or purpose-specific AI models. As such, the safety argument for an AI model within an AI system is captured in either the \textit{Purpose-specific AI Model Safety Argument module} or the \textit{General-Purpose AI Model Safety Argument module}, depending on its intended use (i.e. general or specific purpose). As purpose-specific AI models often build on capabilities provided by general-purpose AI models, an optional link (represented by the hollow circle) can be used to connect the two corresponding argument modules. It is important to note that Retrieval Augmented Generation (RAG) \cite{gao2023retrieval} is increasingly utilised to improve the accuracy of general-purpose AI models for specific purposes \cite{chowdhury2025astrid}. In such cases, the use of RAGs could be justified within the \textit{Purpose-specific AI Model Safety Argument}.

Finally, in this paper, we focus on the structural aspects of the BIG argument. However, a safety case for a complex intervention such as AI is rarely static. Rather, it is a living, dynamic approach integrated into the wider design and operational processes. This approach evolves with new evidence and an updated understanding of the system's actual performance in its intended operational environment \cite{denney2015dynamic}.

The next three subsections explore the argument modules in more detail. Additional practical guidance, and the underpinning assurance methodologies, are detailed in \cite{porter2024principles}\cite{hawkins2022guidance}\cite{hawkins2021guidance}\footnote{Some aspects of PRASE, SACE and AMLAS have been adapted and abstracted to ensure consistency and brevity. Readers are advised to consult the primary sources \cite{porter2024principles}\cite{hawkins2022guidance}\cite{hawkins2021guidance} for a full description of the methodologies and argument patterns.}.

\subsection{AI Safety: The Ethical Argument}
We situate the top-level argument within the wider ethical landscape \cite{morley2020initial}. This is important for three reasons. Firstly, ensuring safety is a fundamental ethical obligation. Secondly, claims about AI safety are inseparable from claims about other ethical principles such as AI fairness and transparency. Thirdly, tradeoffs between various ethical principles (such as transparency and privacy) are often necessary and therefore have to be explicitly considered, justified, challenged and where appropriate accepted.

\begin{figure}
  \centering
  \includegraphics[width=0.6\textwidth]{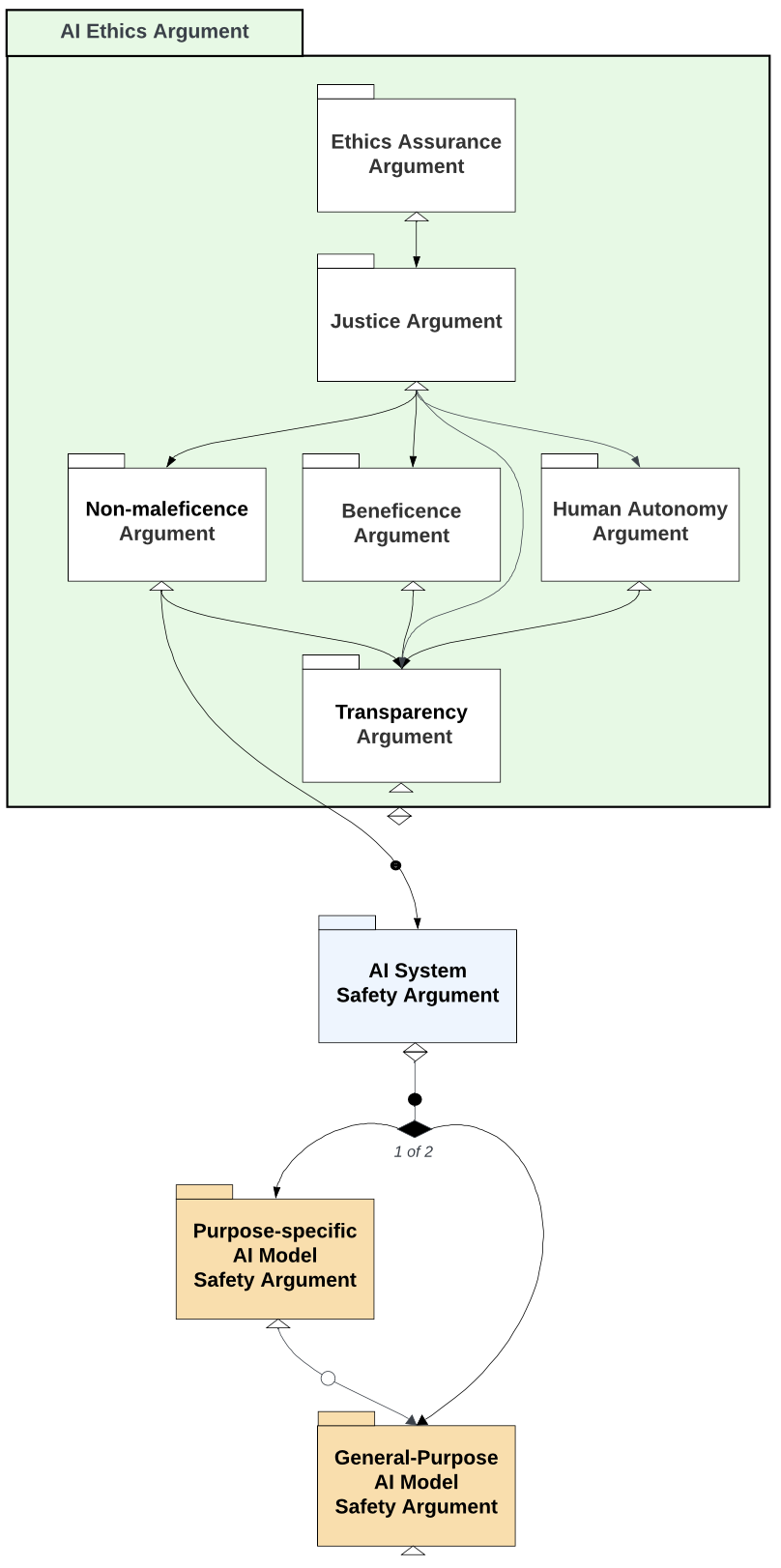}
  \caption{\label{fig:EthicArg}The Ethical Argument represented in GSN}
\end{figure}

Here, we build on the four classical principles of biomedical ethics \cite{beauchamp1994principles}. These are the principles of justice, beneficence (do good), non-maleficence (do not harm) and respect for human autonomy. As we argue in more detail in \cite{porter2024principles}, these principles provide a plausible normative basis and coverage of key ethical values such as sustainability, dignity and reciprocity \cite{kazim2021interrelation}\cite{jobin2019global}\cite{floridi2018ai4people}. Burr and Leslie present an alternative, bottom-up, approach to structuring an AI ethics argument \cite{burr2023ethical}.

The argument for each of the four principles is contained within a separate module in Figure \ref{fig:EthicArg}. The ‘\textit{Ethics Assurance Argument}’ module captures the overall argument for ethical acceptability, which is centred on making the case for the \textit{just} deployment of the AI system. This is detailed in the ‘\textit{Justice Argument}’ module and appeals to the equitable distribution of benefits and harms across all affected stakeholders. This argument deals with the necessary issue of resolving and justifying tradeoffs and builds on and integrates separate and detailed arguments concerning the benefits offered by the use of the AI system (through the ‘\textit{Beneficence Argument}’ module) as well as the mitigation of harm posed by it (via the ‘\textit{Non-maleficence Argument}’ and ‘\textit{Human Autonomy Argument}’ modules).

\begin{figure}
  \centering
  \includegraphics[width=1\textwidth]{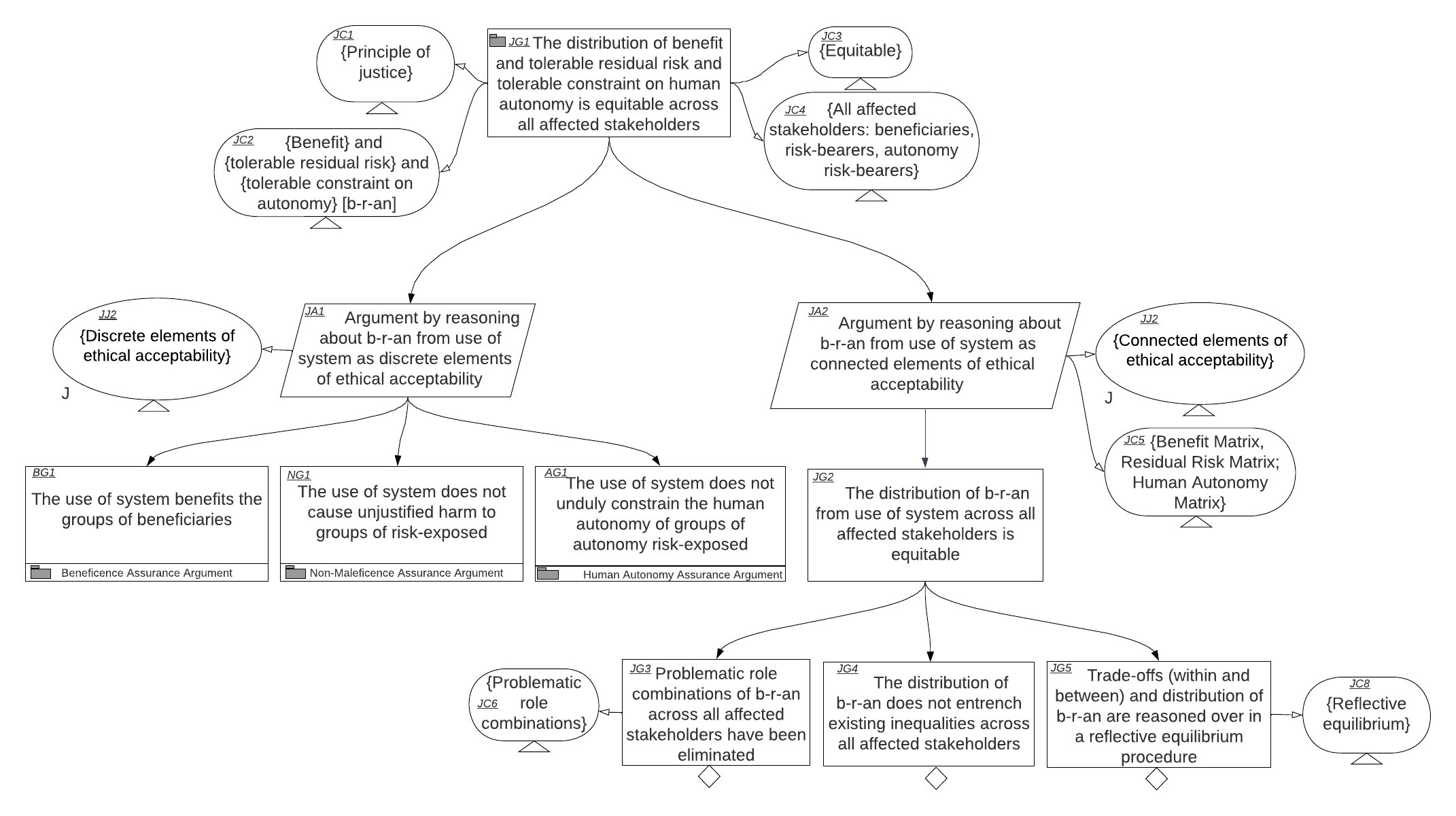}
  \caption{\label{fig:JustArg} Justice argument module of the PRAISE pattern represented in GSN \cite{porter2024principles}}
\end{figure}

It is important to note that the consideration of safety is not limited to the ‘\textit{Non-maleficence Argument}’ but cuts across all argument modules. For example, respecting human autonomy, covered in the ‘\textit{Human Autonomy Argument}’ is fundamental for assuring effective oversight. Otherwise, the role of humans as a risk control in AI-based decision support systems is weakened. This in turn may undermine confidence in the overall safety case. Similarly, reasoning about proportionality, which is central for making decisions about risk acceptability, is considered in the ‘\textit{Justice Argument}’ module, as it often hinges on some form of risk-benefit analysis, including trade-offs.

Another central aspect of the ethical AI debate is transparency \cite{von2021transparency}, captured in the ‘\textit{Transparency Argument}’ module, which is essential for safety assurance. In this argument, we consider transparency as a supporting principle. It plays two key roles in the overall AI safety argument. 
\begin{itemize}
    \item Firstly, it presents transparency claims about the AI development, supply-chain and deployment processes, e.g. why the training datasets were selected and how they were preprocessed to ensure accuracy and balance.
    \item Secondly, it directly links to explainability of the AI outputs and the extent to which the specific formats and modes of explanation are appropriate and meaningful for the intended recipients (e.g. feature importance vs counterfactual reasoning)\cite{miller2019explanation}. We specifically build on the philosopher Paul Grice’s maxims of cooperative communication \cite{grice1975logic}\cite{kaas2024assuring}. The four maxims of quantity, quality, relevance and manner provide a sound basis for assuring the suitability and effectiveness of communication about critical AI properties.
\end{itemize}

The \textit{AI Ethics Argument} is described in full in \cite{porter2024principles}. However, to illustrate how this argument brings the different ethical considerations together, we next discuss the\textit{ Justice Argument} in more detail. The argument contained within this module is shown in Figure \ref{fig:JustArg}. The top claim states that the “\textit{distribution of benefit, tolerable residual risk, and tolerable constraint on human autonomy is equitable across all affected stakeholders}”. 

This claim appeals to the notion of distributive justice, reflecting a deeper consideration of who benefits and who bears the risks from the use of AI. This in turn provides a more transparent basis for judging the proportionality of risks to different stakeholders and at what cost or benefit to others and themselves \cite{porter2024principles}.

To support this claim, the argument is built on two overarching strategies. The first, \textit{JA1}, details claims that the system provides benefits, does not cause unjustified harm and does not unduly place constraints on human autonomy. This strategy considers these issues as discrete elements of ethical acceptability, where each is developed further in a separate argument module. The second strategy, \textit{JA2}, collectively considers benefits, risks of harm and constraints on human autonomy, focusing on inevitable tradeoffs and their justification (JG5), e.g. through reflective equilibrium \cite{cath2016reflective}, associated with John Rawls' work on justice \cite{rawls2017theory}. Two additional claims are emphasised in this argument: that unacceptable role combinations are eliminated (\textit{JG3}) e.g. certain groups only bearing risk, and no benefit, from AI’s use, and that AI deployment does not entrench existing inequalities (\textit{JG4}).

A key challenge in supporting the \textit{Ethical Argument} lies in the limited availability of practical methods and techniques for translating ethical principles into concrete requirements \cite{morley2020initial}. Example \ref{exp:RAD} illustrates a potential approach to addressing this challenge.


\begin{example}	[Identifying Ethical Concerns for an AI-enabled Assisted Dressing Robot]
\label{exp:RAD}
AI-enabled robotics have been proposed as a way of improving the lives of those living long-term conditions which restrict physical capabilities. Deploying such systems could increase independence in the elderly and reduce the need for traditional care services. The services offered by such systems will need to be personalised to the user and the context in which they are to be deployed. 

\begin{center}
    \includegraphics[width=0.4\textwidth]{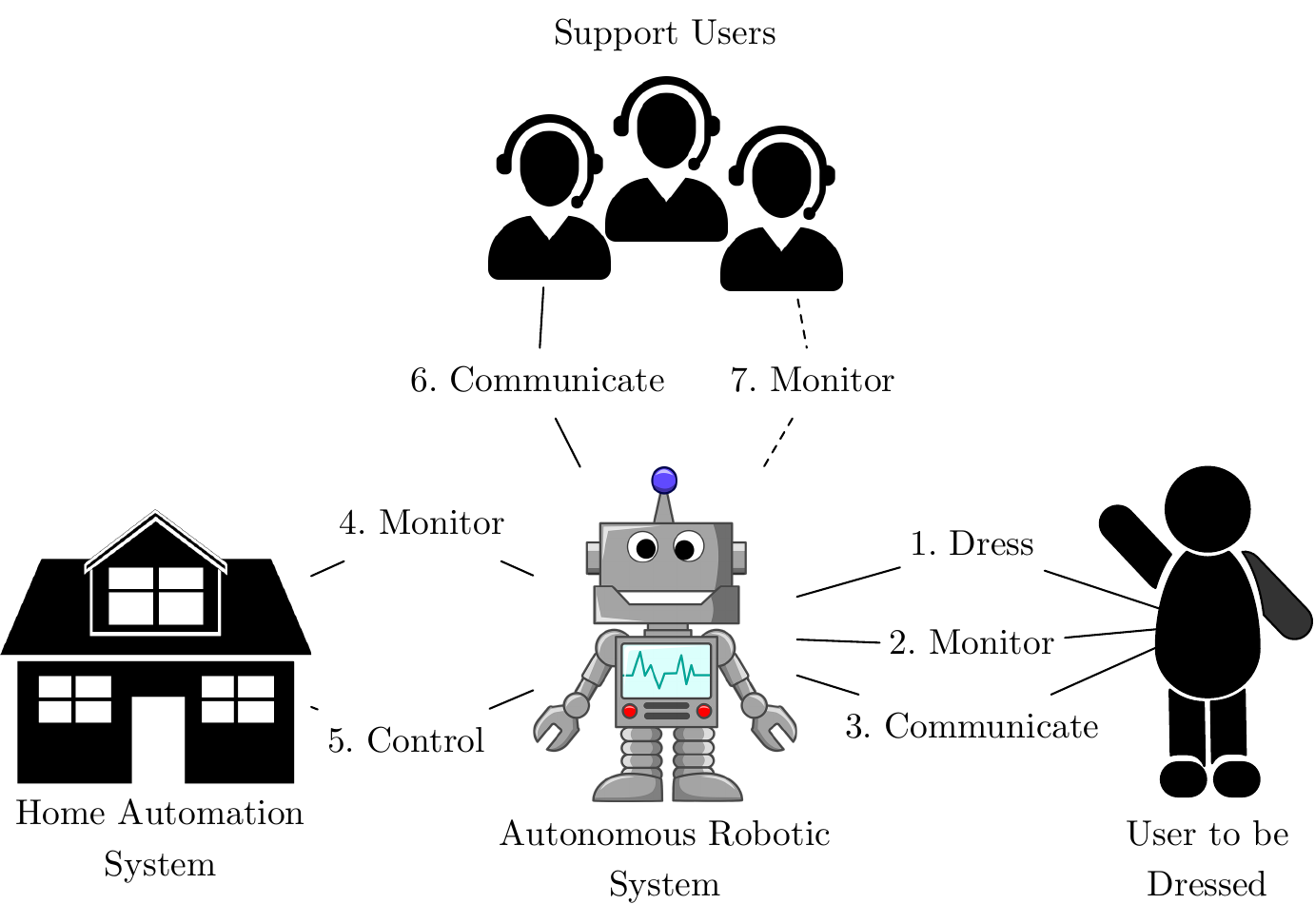}
\captionof{figure}{\label{fig:RAD}Robotic assisted dressing platform \cite{townsend2022pluralistic}}
\end{center}

Figure~\ref{fig:RAD} shows one such case where an AI-enabled robotic system is deployed to aid a user to dress in their own home \cite{townsend2022pluralistic}\cite{jevtic2018personalized}. In order to undertake the primary function of dressing, the platform will need to monitor the user and communicate with them. Since dressing necessitates the user to be undressed (or partially dressed) the platform is also able to control the heating, curtains and lighting in the home. Finally, since the user may be vulnerable, facilities are available to communicate with a support team who, in turn, can monitor the state of the system to ensure it is functioning as expected.

For critical systems such as this, it is necessary for us to derive requirements which are not only functional in nature, but also respect Social, Legal, Ethical, Empathetic and Cultural (SLEEC) norms \cite{townsend2022pluralistic}. These norms are derived from high-level principles (Table~\ref{tab:sleec}) and refined through a structured elicitation process (Figure~\ref{fig:SLEEC_process}) to define rules and address the trade-offs arising from the context into which the system is to be deployed and the multi-disciplinary requirements of system stakeholders. 

\vspace{3mm}
    \begin {sffamily}
    \begin{footnotesize}
    \begin{tabular}{lp{9.6cm}}
    \toprule
         \textbf{SLEEC Concern} & \textbf{Description}  \\ \midrule
         Privacy & Limiting intrusion on the personal space of the user and ensuring privacy is protected; safeguarding health data, practising good data stewardship, and granting or restricting access to medical records\\
         Respect for Autonomy & Granting and withdrawing of permissions, including consent and assent; ensuring the user maintains an appropriate level of control\\
         Dignity & Understanding and accommodating the user's social and cultural sensitivities, respectful treatment\\
         Explainability and transparency & Informing the user about system decision-making and any inferences made; providing justification for a course of action adopted\\ 
         Beneficence & Maximising good outcomes \\
         Non-maleficence & Minimising harm by ensuring safety and reducing the possibility of physical and psychological harm to the user\\\bottomrule
    \end{tabular}
    \end{footnotesize}
    \end {sffamily}
    \captionof{table}{\label{tab:sleec} Examples of SLEEC concerns in the robotic asssisted dressing system \cite{townsend2022pluralistic}}

The complexity of the rules arising from such systems could lead to conflicts and redundancy. The SLEEC methodology may support the \textit{AI Ethics Argument} by providing a potential approach to identifying and analysing key ethical issues and how they relate to other social norms. It offers a language, formal semantics and a toolset to encode these rules for use in robotic and AI systems and part of the evidence base needed to support the top argument.

\begin{center}
    \includegraphics[width=0.6\textwidth]{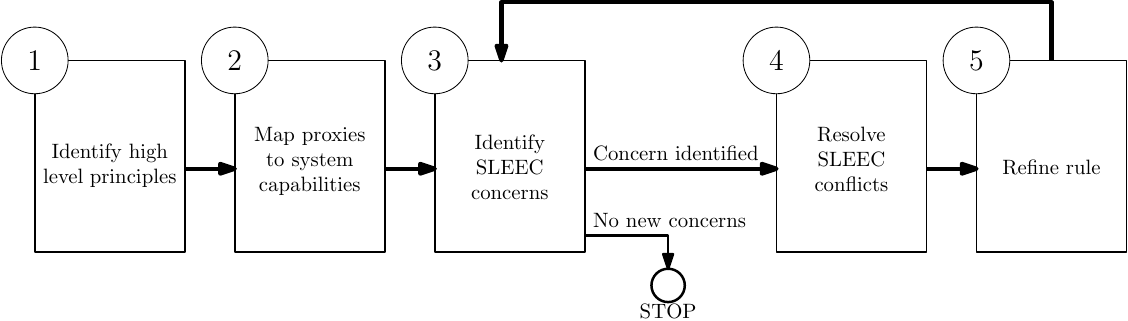}
\captionof{figure}{\label{fig:SLEEC_process}The SLEEC requirements elicitation process \cite{townsend2022pluralistic}}

\end{center}
\end{example}

\section{AI Safety: The System Argument}
This argument refines the ethical claims about AI safety, considering them in more detail within both the context of the wider system (e.g. a maritime autonomous surface ship \cite{MASS_VOJKOVIC2020333}\cite{MASS_GOERLANDT2020104758}) and the social and organisational setting in which it is deployed (e.g. passenger ferry services in busy maritime environments and staffing levels that impact on the ability of people to augment autonomous AI functions).

Issues of particular focus are the scope of the operational domain, considering the system’s ability to safely operate within and beyond its defined context, and human-machine interactions, including challenges around over- or under-reliance on AI-enabled functions. The systems engineering perspective becomes more prominent here, focusing on claims and assumptions about the overall system architecture, including redundancy, diversity and monitoring (see Example \ref{exp:Arch}: architectural tactics for incorporating AI in aviation). It also addresses how detailed safety requirements, incorporating uncertainties, are determined, refined, and verified.

\begin{example} [Architecting AI-based Autonomous Aviation Systems]
\label{exp:Arch}
When using AI in safety-critical systems it is crucial to take a systems perspective to understand and manage its contribution to risk. Architectural patterns can be used to mitigate some AI failure modes or uncertain behaviours. For example, it may be possible to monitor and constrain outputs from an AI-based drone flight stabilisation component to prevent unexpectedly large directional changes being sent to propellers. Such outputs could physically strain the drone and/or lead to sudden and unpredictable trajectories around infrastructure \cite{ryan2024bridging}. A monitor-based architecture thus reduces the AI component's individual contribution to risk of collision and improves reliability of the overall system. Further, it may make assurance requirements on the AI model safety argument less onerous.

There are many different architectural design patterns (e.g. component monitoring,  component switches, voting on outputs from multiple diverse components with the same function) which can be combined to help incorporate AI into avionics systems and maintain existing high-assurance norms \cite{fenn2023architecting}.  For example, runtime monitors can capture information on real-time performance of AI components, switching to a traditional (but perhaps less adaptive) alternative function if the performance is below a particular threshold (Figure \ref{fig:ArchPattern2})\cite{machin2016smof}. These patterns of architectural designs have been used for avionics systems for many years, but AI provides additional challenges to their efficacy. For example, run-time performance of an image classifier is typically difficult to accurately assess due to the lack of ground truth for comparison. 

\vspace{3mm}
\begin{center}
    \includegraphics[width=0.5\textwidth]{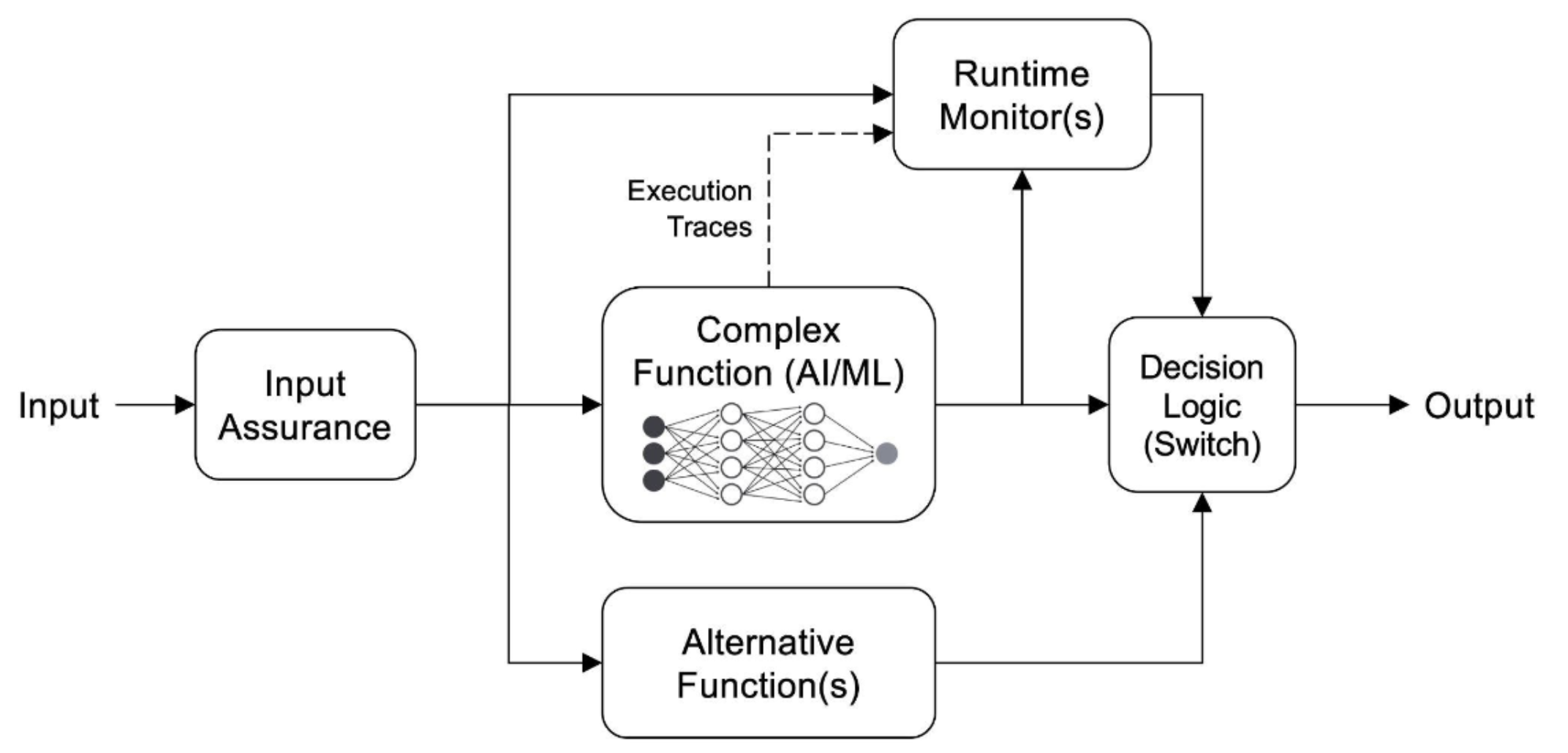}
\captionof{figure}{\label{fig:ArchPattern2}Runtime assurance (RTA) architectural pattern \cite{fenn2023architecting}}
\end{center}

\end{example}

\begin{figure}
  \centering
  \includegraphics[width=0.75\textwidth]{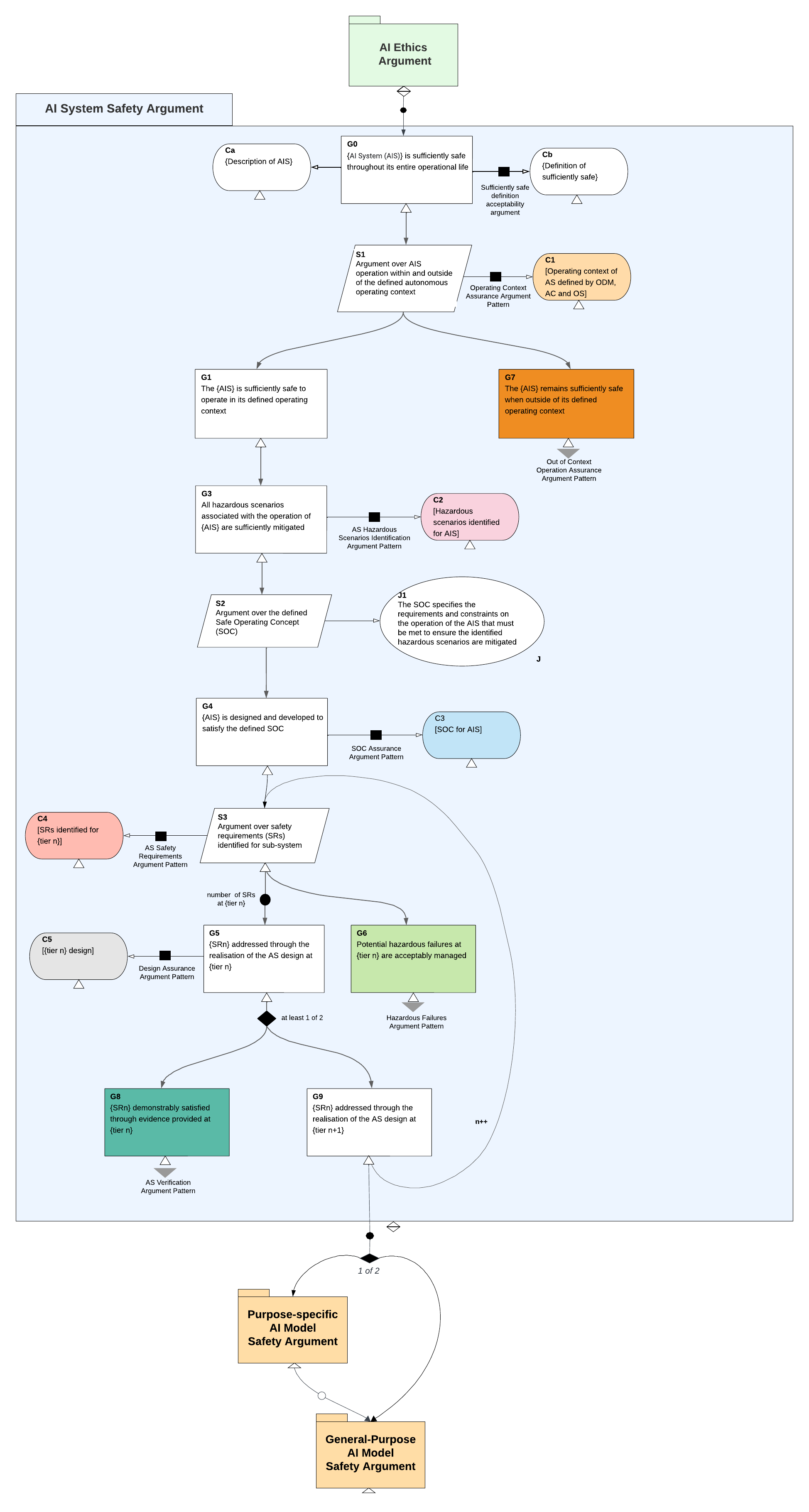}
  \caption{\label{fig:SysArg} The AI System Argument represented in GSN}
\end{figure}

Figure \ref{fig:SysArg} depicts the ‘\textit{AI System Safety Argument}’ module. The top-level claim (\textit{G0}) is that the “\textit{\{AI System (AIS)\} is sufficiently safe throughout its entire operational life}”, with the curly brackets indicating that the term ‘\textit{AI System (AIS)}’ requires instantiation. The argument strategy supporting this claim centres on the ability of the system to remain sufficiently safe within (\textit{G1}) and outside (\textit{G7}) its defined operating context (see Example \ref{exp:FRAM}: modelling intravenous infusion administration in intensive care units). Following that, the argument takes a hazard-based approach (\textit{G3}), focusing on mitigating the identified hazardous scenarios by developing safety requirements and constraints on the operation of the system. Collectively, these requirements and constraints constitute the Safe Operating Concept (SOC) (\textit{J1}). The argument is iterative in nature, decomposing assurance claims about the development and refinement of the safety requirements at different levels of abstraction, as indicated in \textit{G9} and \textit{S3} (see Example \ref{exp:Space}: generating AI safety requirements for satellite-based wildfire detection).

\begin{example} [Complex Contexts: Performance Variability in Intravenous Infusion Administration in ICU]
\label{exp:FRAM}
The use of AI in healthcare is high on the political agenda. However, to meaningfully incorporate AI-based tasks into clinical pathways, it is important to model and analyse existing clinical needs, challenges and constraints. Appreciating the complexity of clinical practice and reducing the gap between work-as-done and work-as-imagined is essential \cite{hollnagel2018safety}. FRAM (Functional Resonance Analysis Method) is a well-established technique in safety science used to model the performance variability of complex sociotechnical systems (work-as-done) \cite{hollnagel2017fram}.

Figure~\ref{fig:FRAM} shows a FRAM diagram of intravenous infusion administration \cite{furniss2020using}. This links to the need to define the operating context of the system (Context \textit{C1} in the \textit{AI System Argument}). The purpose behind the model and analysis, detailed here \cite{furniss2020using}, was to ensure sufficient understanding of current practice as a prerequisite for automating any tasks using AI systems.

\vspace{3mm}
\begin{center}
    \includegraphics[width=0.9\textwidth]{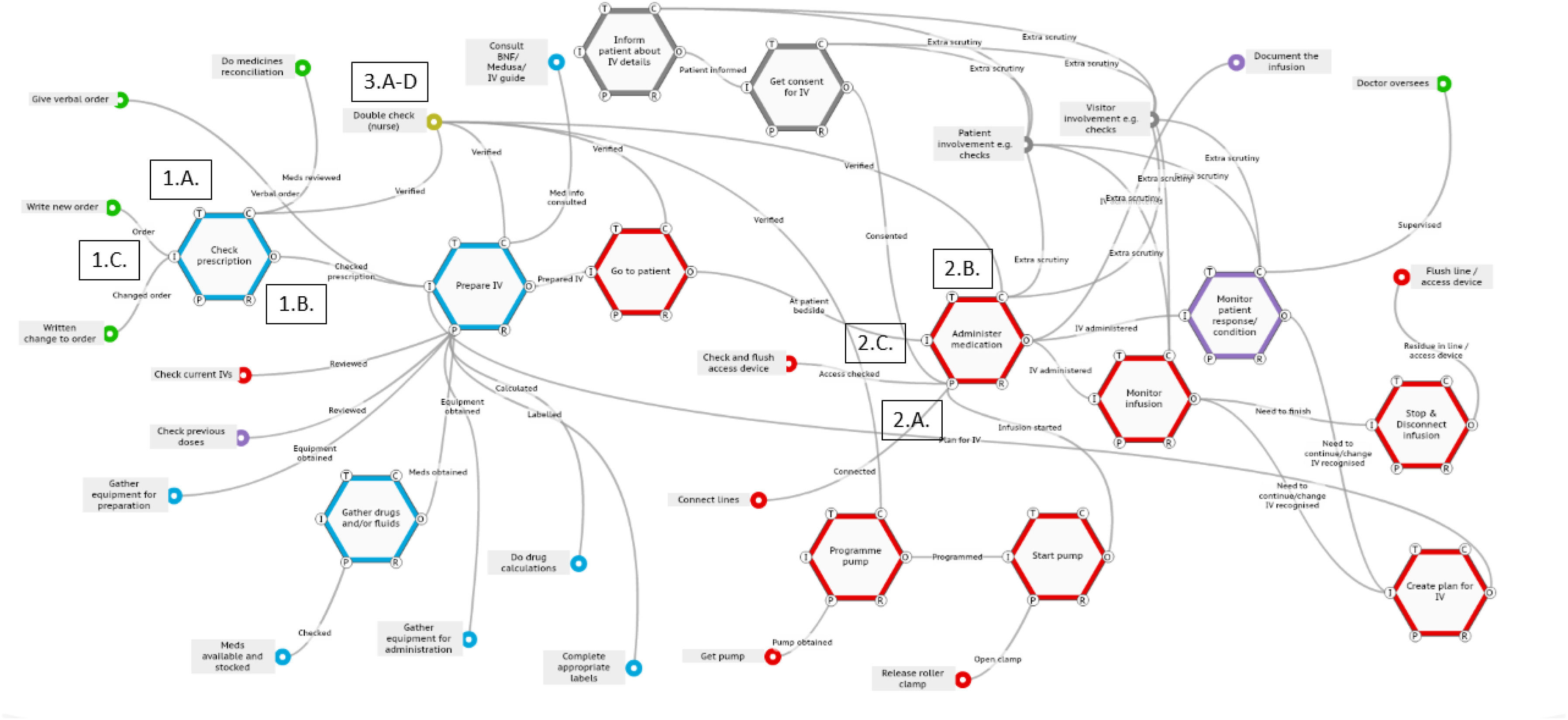}
\captionof{figure}{\label{fig:FRAM}FRAM network diagram of intravenous infusion administration \cite{furniss2020using}}

\end{center}

For example, the model indicates that some variability in medication ordering in ICU may be justified. The standard operating procedure (SOP) suggests nurses should always have a written prescription beforehand to ensure correct administration. However, emergencies may require immediate drug administration, or doctors may be too busy to write an order immediately and advise proceeding without it, on the understanding that the prescription will be issued later. This illustrates how people make adjustments in everyday work in order to deliver care successfully given competing demands and priorities.  When designing and deploying AI-based systems, it is important to consider whether and how their use might potentially disrupt people's ability to make such adjustments flexibly, e.g. if the AI requires that a prescription has been issued on the electronic system without the ability to afford any flexibility.  

\end{example}

It is important to note that Assurance Claim Points (ACPs), represented as black squares, are used in the representation of this argument \cite{hawkins2011new}. ACPs provide links to confidence arguments that justify the sufficiency of confidence in specific aspects of the safety argument. For example, for a hazard-based argument such as the one presented in Figure \ref{fig:SysArg},  confidence that all hazardous scenarios associated with the operation of the system are identified is fundamental. To this end, an ACP is added to the contextual link between \textit{G3} and \textit{C2}, creating a pointer to a detailed confidence argument concerning the way in which scenarios were generated, reviewed and updated. The full argument is explained in detail in \cite{hawkins2022guidance}.

\begin{example} [AI Safety Requirements for Satellite-Based Wildfire Detection and Alert System]
\label{exp:Space}
A satellite with a multi-spectral imager passes over a region of interest that may contain wildfires (Figure \ref{fig:SATOverview1}). An Artificial Neural Network (ANN) onboard the satellite is trained to detect wildfires in the images received and to send an alert to a ground station identifying the location of the fire \cite{hawkins2023creating}. This alert can then be passed to the relevant authorities who can respond appropriately. Detecting wildfires onboard a satellite reduces bottlenecks and delays associated with sending image data to be processed on the ground \cite{barmpoutis2020review}.

\begin{center}
    \includegraphics[width=0.5\textwidth]{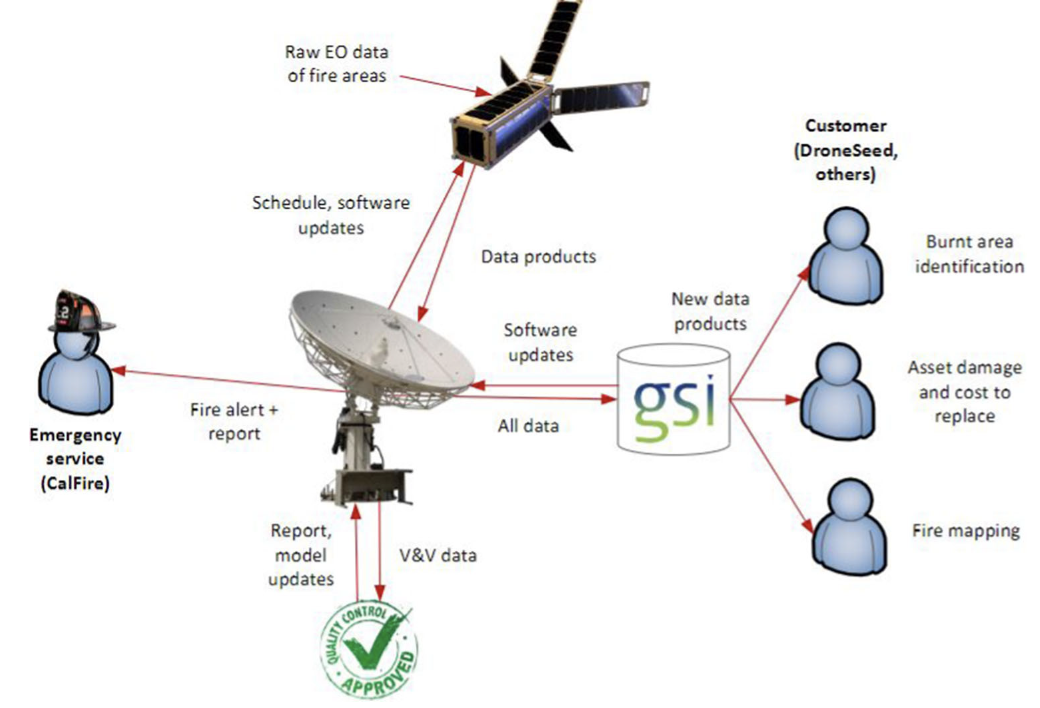}
\captionof{figure}{\label{fig:SATOverview1}Concept of Operations for Wildfire Alert System \cite{hawkins2023creating}}
\end{center}

Two potential hazards were identified for the wildfire alert system:
\begin{itemize}
    \item \textbf{Hazard 1} is that the emergency services miss a wildfire emergency, which could lead to a delay in the response to the fire, a larger and less controlled fire, and thus a potential increase in the risk of harm to people and property or putting firefighting teams in danger. It is determined that the AI-based wildfire alert system could contribute to this hazard through failure to detect the presence of a wildfire. Table \ref{fig:SatRQ1} shows three safety requirements identified for the wildfire alert system in order to mitigate the contribution of the system to this hazard through the specification of required accuracy and response time of the AI-based wildfire alert system.
    \item \textbf{Hazard 2} is that an alert is raised for a wildfire that does not actually exist. This could result in fire response resources being mis-assigned and thus unavailable to respond to real wildfires in a timely manner. The AI-based wildfire alert system could contribute to this hazard through detecting a wildfire in the incorrect location. Figure \ref{fig:SatRQ1} shows an associated safety requirement for the wildfire alert system that specifies an acceptable rate of false detections through comparison to current systems.
\end{itemize}

    \begin{center}
    \begin {sffamily}
    \begin{footnotesize}
    \captionof{table}{\label{fig:SatRQ1}System safety requirements for wildfire alert system \cite{hawkins2023creating}}\vspace{-2mm}
    \begin{tabular}{lp{10.6cm}}
    \toprule
    \multicolumn{2}{l}{\textbf{Hazard 1 - Services Miss an Emergency}}\\
    \textbf{REQ-SAFE-ER-1} & The Emergency Response Service shall determine the location of an active wildfire within 200m of its true location.\\
     \textbf{REQ-SAFE-ER-2} & The Emergency Response Service shall inform emergency services of an active wildfire within 3 hours of it starting.\\
     \textbf{REQ-SAFE-ER-3} & The Emergency Response Service shall positively identify 95\% of all active wildfires acquired by the satellite instrument within the area of interest.\\
     \midrule
      \multicolumn{2}{l}{\textbf{Hazard 2 - Services are Directed to a False Emergency}}\\
    \textbf{REQ-SAFE-ER-4} & The Emergency Response Service shall falsely indicate active wildfires in the area of interest at a rate not exceeding current fire alert service (average for FIRMS of 52 per month) .\\\bottomrule
    \end{tabular}
    \end{footnotesize}
    \end{sffamily}
    \end{center}

This safety requirement specification provides Context \textit{C4} to the argument in Figure \ref{fig:SysArg}, and justification for the sufficiency of these safety requirements is provided as a confidence argument.
\end{example}

\section{AI Safety: The Purpose-specific AI Model Safety Argument}
This argument starts to cover the technical aspects of AI safety. In particular, we focus on Machine Learning (ML) in its different forms, such as supervised, unsupervised, and reinforcement learning. The argument considers ML-based functions deployed to serve a specific and often narrow purpose, e.g. diagnosis of particular clinical diseases in specific pathways. The safety claims, assumptions and evidence in this argument cover the entire ML lifecycle, including data curation, model training and testing, and subsequent deployment, monitoring and updates (following the AMLAS methodology \cite{hawkins2021guidance}). The argument is largely requirements-driven, justifying how system-level safety requirements, considered in the \textit{AI System  Argument} above, are broken down into specific technical AI safety requirements. Evidence that these requirements have been validated and verified is then generated within the wider system and environment. Key claims centre on specific performance and robustness metrics, quantified safety thresholds, accuracy and representativeness of the datasets and explainability of the model outputs. Fundamental concerns include \textit{traceability} between the technical claims at this level and the higher-level safety claims at the system and ethical/societal levels.

\begin{table}
	\centering
	\caption{Assurance methods for the Data Management stage adapted from~\cite{ashmore_assuring_2021}}
	\begin{footnotesize}
	
	\vspace*{-2mm}
	\def\tabcolsep{1.4pt}
	\sffamily
	\begin{tabular}{L{4.25cm}cccccccc} 
	\toprule
        & \multicolumn{4}{c}{\textbf{Associated activities$^\dagger$}} & \multicolumn{4}{c}{\textbf{Supported desiderata$^\ddagger$}}\\ \cmidrule(l{1pt}r{2pt}){2-5}\cmidrule(l{2pt}r{1pt}){6-9}
		\textbf{Method} & \hspace*{-0.3mm}Collection & Preprocess. & Augment. & Analysis & Relevant & Complete & Balanced & Accurate \\ \midrule
		\\[-1.5em]
   	    \rowcolor{gray!25}
		Use trusted data sources, with data-transit integrity guarantees & \ding{52} & & & & \ding{72} & & & \\ 
		Experimental design & \ding{52} & & \ding{52} & & \ding{72} & \ding{72} & \ding{73} & \\ 
   	    \rowcolor{gray!25}
		Simulation verification and validation  & & & \ding{52} & & \ding{72} & \ding{73} & \ding{73} & \\ 
		Exploratory data analysis & & & & \ding{52} & & \ding{72} & \ding{72} & \\ 
   	    \rowcolor{gray!25}
		Use adversarial examples & & & \ding{52} & & \ding{73} & \ding{72} & & \\ 
		Include a ``dustbin'' class  & & & \ding{52} & & \ding{73} & \ding{72} & & \\ 
   	    \rowcolor{gray!25}
		Remove unwanted bias & & \ding{52} & \ding{52} & & \ding{72} & & \ding{73} & \\
		Compare sampling density  & & & \smltick & \ding{52} & & \ding{72} & \ding{73} & \\ 
   	    \rowcolor{gray!25}
		Identify empty and single-class regions  & & & \smltick & \ding{52} & & \ding{72} & \ding{73} & \\ 
		Use situation coverage & & & & \ding{52} & & \ding{72} & & \\ 
   	    \rowcolor{gray!25}
		Examine system failure cases & & & & \ding{52} & & \ding{72} & & \\ 
		Oversampling \hspace*{-0.2mm}\&\hspace*{-0.2mm} undersampling  & & & & \ding{52} & & \ding{72} & \ding{72} & \\ 
   	    \rowcolor{gray!25}
		Check for within-class and feature imbalance & & & & \ding{52} & & \ding{72} & & \\ 
		Use a GAN & & & \ding{52} & & & \ding{72} & \ding{73} & \\ 
   	    \rowcolor{gray!25}
		Augment data to account for sensor errors & \smltick & & \ding{52} & & \ding{73} & & & \ding{72} \\
		Confirm correct software behaviour & \smltick & \ding{52} & \ding{52} & \smltick & \ding{73} & \ding{72} & \ding{73} & \ding{73} \\ 
   	    \rowcolor{gray!25}
		Use documented processes & \ding{52} & \ding{52} & \ding{52} & \ding{52} & \ding{73} & & & \ding{72} \\ 
		Apply configuration management  & \ding{52} & \ding{52} & \ding{52} & \ding{52} & \ding{73} & & & \ding{72} \\[-0.5mm] \bottomrule
		\multicolumn{9}{l}{$^\dagger$\ding{52} = activity that the method is typically used in; \smltick = activity that may use the method}\\
		\multicolumn{9}{l}{$^\ddagger$\ding{72} = desideratum supported by the method; \ding{73} = desideratum partly supported by the method}
	\end{tabular}
	\end{footnotesize}
	\label{tab:dm_astech}
\end{table}

\begin{figure}
  \centering
  \includegraphics[width=1\textwidth]{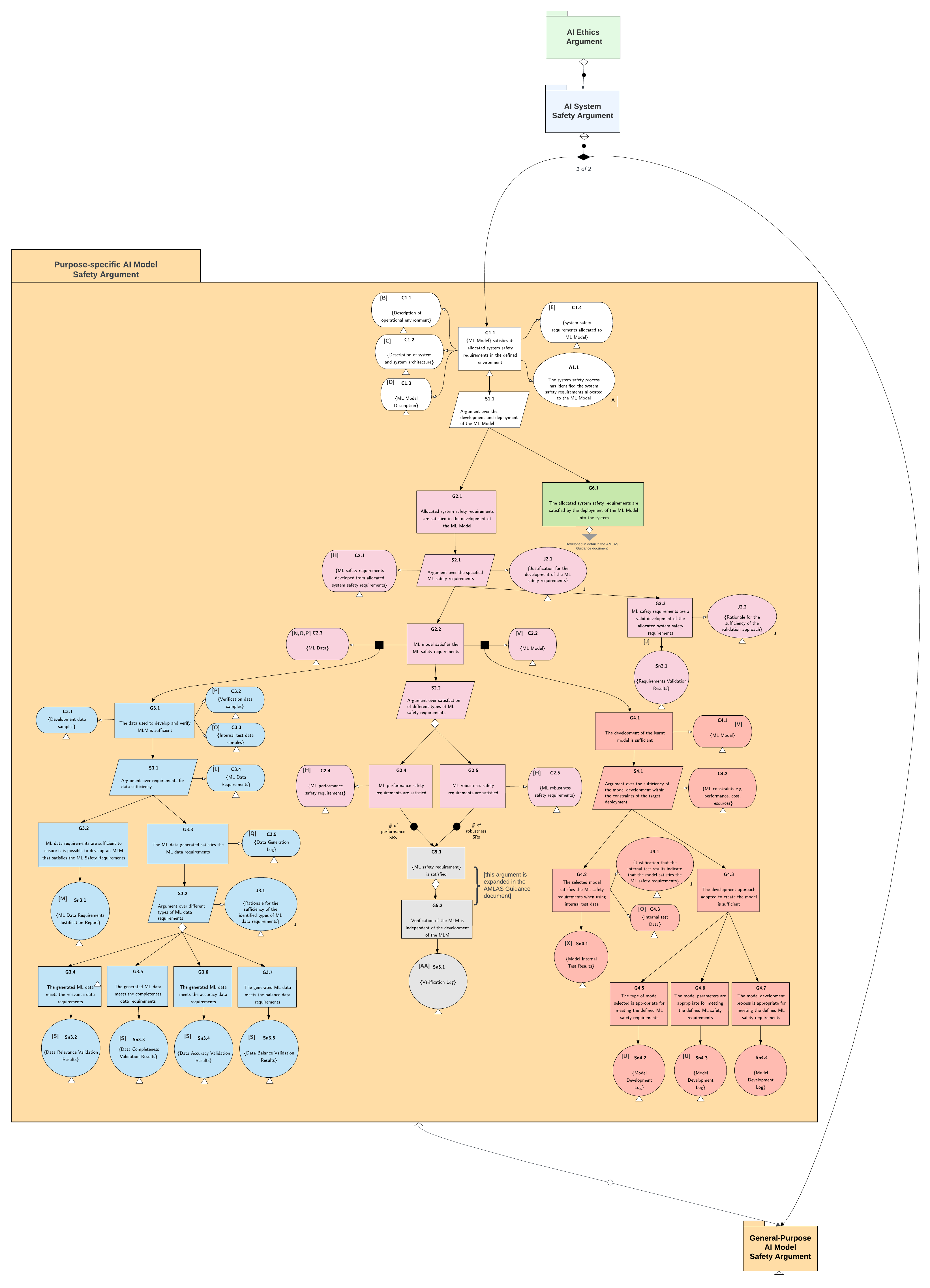}
  \caption{\label{fig:MLArg}The Purpose-specific AI Model Safety Argument represented in GSN (bird's-eye view)}
\end{figure}

Here, we build on the AMLAS methodology to structure the pattern for a purpose-specific AI model safety argument. Figure \ref{fig:MLArg} depicts a simplified and adapted composition of the 6 sub-arguments comprising the AMLAS safety argument pattern. These correspond to safety assurance within the following interrelated stages in the ML lifecycle \cite{hawkins2021guidance}:

\begin{enumerate}
    \item ML Safety Assurance Scoping
    \item ML Safety Requirements Assurance
    \item Data Management Assurance
    \item Model Learning Assurance
    \item Model Verification Assurance
    \item Model Deployment Assurance
\end{enumerate}

\begin{example} [Safety-driven Design of Machine Learning for Sepsis Treatment]
\label{exp:Sepsis}
Sepsis, a life-threatening organ dysfunction caused by a dysregulated host response to infection, stands as one of the leading causes of mortality and one of the most resource-intensive conditions to treat in hospitals. The use of Reinforcement Learning (RL) can help to discover an optimal treatment strategy, particularly optimising the administration of vasopressors and fluids, which are two fundamental medications for sepsis treatment.

Whilst learning the optimal treatment, the RL system also must not learn hazardous behaviours. One of the hazardous scenarios is a sudden change of vasopressor dose, which can cause significant harm to the patients, e.g. resulting in acute hypotension (arising from rapidly decreasing doses), hypertension or cardiac arrhythmias (arising from rapidly increasing doses) \cite{jia2021safety}. Therefore, we evaluated whether such behaviours exhibited in the original learnt policy, showing that 35\% of the cases that the RL model would recommend sudden change in vasopressor dose compared to 3\% in clinician policy, i.e. what clinicians have done for the same patient cases, as shown in Table~\ref{tab:dose}.

\vspace{3mm}
\centering
\captionof{table}{\label{tab:dose}Summary of max dose change between consecutive doses for the three policies}
\label{tab:Comp-table}
\begin{tabular}{|l|l|l|}
\hline
                 & \multicolumn{2}{l|}{Dose of vasopressor (mcg/kg/min)} \\ \hline
 & \begin{tabular}[c]{@{}l@{}}Small-Medium Dose Change (0-0.75)\end{tabular} & \begin{tabular}[c]{@{}l@{}}Large Dose Change (\textgreater{}0.75)\end{tabular} \\ \hline
Clinician Policy & 97\% (2,100)               & 3\% (60)                 \\ \hline
Original Policy  & 65\% (1,404)               & 35\% (756)               \\ \hline
Modified Policy  & 92\% (1,990)               & 8\% (170)                \\ \hline
\end{tabular}
\vspace{3mm}

Guided by AMLAS, especially the model learning stage, we modified the loss function and the feature space in the RL model, as shown in Table~\ref{tab:changes}, then retrained the RL model. The resulting modified policy showed only 8\% of sudden vasopressor dose change when evaluated on the same patient cases, which is much closer to the clinicians’ behaviour. This shows the value of following a systematic approach for proactively mitigating the system level risk in the context of the machine learning lifecycle. 

\vspace{3mm}
\captionof{table}{\label{tab:changes}Major changes in the modified RL model}
\label{tab:my-table}
\resizebox{\textwidth}{!}{%
\begin{tabular}{|l|l|l|}
\hline
 & \textbf{Features in state space (R1)} & \textbf{Cost Function(R3)} \\ \hline
\textbf{Original RL model} & 48 & \begin{tabular}[c]{@{}l@{}}$ L(\theta) = E[(Q_{double-target} - Q(s,a;\theta))^2] $ +\\ $ \lambda_1 max(|Q(s,a;\theta)| - Q_{thresh}, 0) $  \end{tabular} \\  \hline
\textbf{Modified RL model} & \begin{tabular}[c]{@{}l@{}}48 (Removed one feature\\ -- timestep,  added an extra\\ one -- relative dose change )\end{tabular} & \begin{tabular}[c]{@{}l@{}}$ L(\theta) = E[(Q_{double-target} - Q(s,a;\theta))^2] $ +\\ $ \lambda_1 max(|Q(s,a;\theta)| - Q_{thresh}, 0) $ + \\ $ \lambda_2 max(|V_{change}|- 0.75, 0) $ \\ $V_{change}$ is the agent recommended dose (argmax \\ of $Q(s,a; \theta)$) minus the vasopressor dose \\ in the previous step; $\lambda_1$ and $\lambda_2$ are the tuning \\ parameters that decide how much to penalise the \\ flexibility of the model.
\end{tabular} \\ \hline
\end{tabular}%
}
\vspace{3mm}

\end{example}

The focus on technical development considers ML developers, systems engineers and subject matter experts (e.g. clinicians or pilots) as the primary stakeholders in generating the evidence necessary at this stage. The process outlined in the AMLAS methodology connects the SACE and AMLAS patterns through the safety assurance scoping argument, which maps the role of the ML model under consideration to the potential for hazardous scenarios that may arise from the functions undertaken by the ML model. As such, the top-level claim (\textit{G3.1}) made by the application of AMLAS is that the “\textit{{ML Model} satisfies its allocated system safety requirements in the defined environment}”. See Example \ref{exp:Sepsis}, illustrating a traceable link between clinical hazard and the training requirements for an RL agent involved in supporting the treatment of sepsis.

Further, given the data-intensive nature of ML, the argument pattern pays particular attention to the justification of the choice of the training and verification datasets. For instance, the safety claim (\textit{G3.2}) creates an assurance link between the ML safety requirements (traced to system-level safety requirements) and the data requirements. That is, the data requirements are sufficient for realising the system-level safety considerations at the data level. This is then refined to consider the rationale for specific data desiderata, primarily relevance, completeness, accuracy, and balance (see Example \ref{exp:Rare}, illustrating an approach to mitigating the impact of rare subclasses in deep neural network classifiers).

\begin{example} [Detection and Mitigation of Rare Subclasses in Deep Neural Network Classifiers]
\label{exp:Rare}
Legal frameworks make explicit lists of protected characteristics which allow for us to define measures of fairness against which we can evaluate our systems. These characteristics can be used in the development of data collection activities as well as evaluation processes, e.g. ensuring that gender bias is avoided.

Unfortunately these lists are often too coarse to identify pockets of intersectional data where individuals may still be unfairly treated in practice. Monitoring and mitigating such subclass discrimination requires us to identify rarity within the data used to train our models and observed at run-time \cite{paterson2021detection}.

Figure \ref{fig:rare} illustrates a process in which data samples are evaluated with a simple commonality score that is correlated to the probability of misclassification, and hence unfair bias, in the resulting AI-enabled system. Samples which are dissimilar to the greater population are then examined to identify common features and mitigations (such as data augmentation of collection) is undertaken to correct for the bias. 

A similar assessment of the commonality score at run-time allows us to present this to the user, along with our prediction to allow for adjustments to be made at run time. This method represents an example approach for justifying the quality and coverage of the AI data management process.

\vspace{3mm}
\begin{center}
    \includegraphics[width=0.6\textwidth]{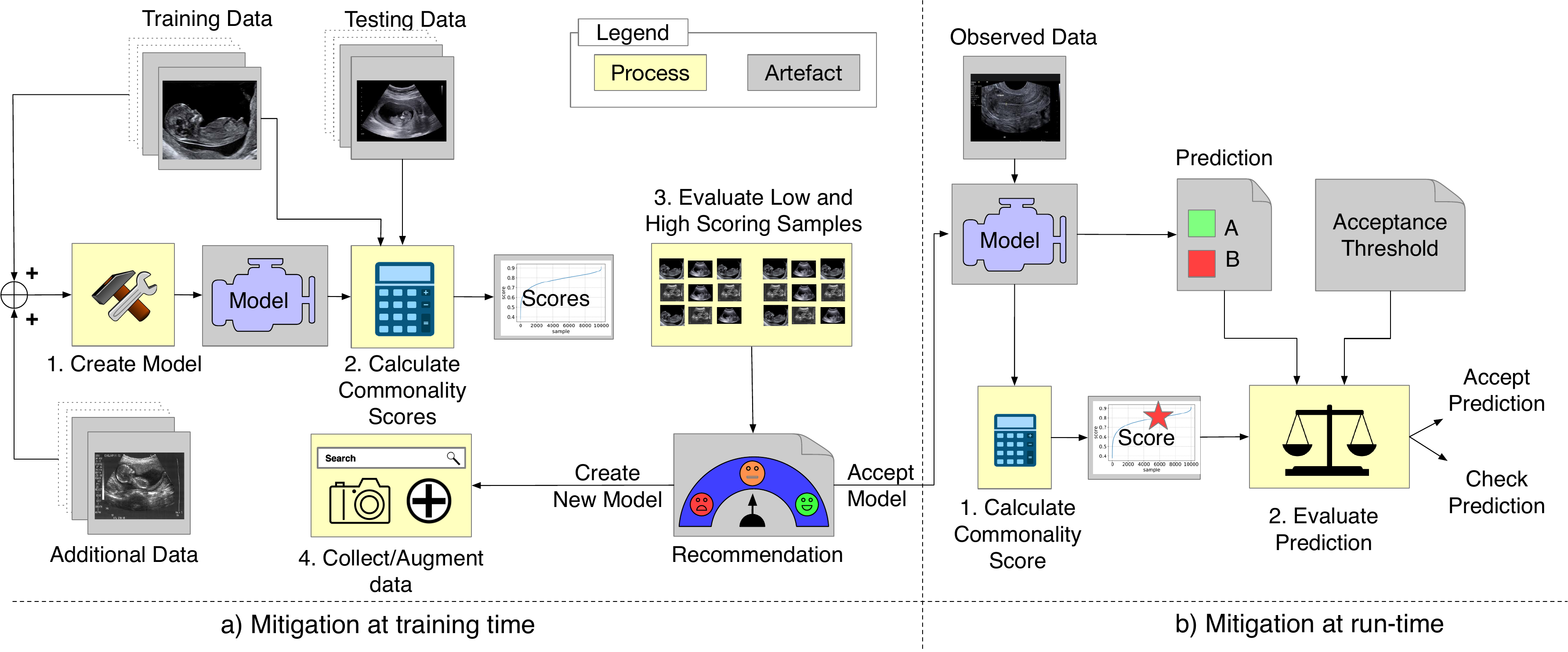}
\captionof{figure}{\label{fig:rare}Processes for mitigating rare subclasses at training time (left) and run-time (right). Adapted from~\cite{paterson2021detection}}
\end{center}

\end{example}

No single tool or technique can provide comprehensive safety evidence across all stages of the ML lifecycle. However, a systematic approach can use sets of diverse and complementary tools and techniques to create a compelling safety argument for each stage. Table~\ref{tab:dm_astech} shows how considering tasks involved with data management and the desirable features of data can build evidence \cite{ashmore_assuring_2021}. Such evidence supports the claim that the data used to develop and verify an ML component is sufficient (Example \ref{exp:DeepCert} provides an example technique for generating such verification evidence).

\begin{example} [Verification of Contextually Relevant Robustness for Neural Network Image Classifiers]
\label{exp:DeepCert}
Traditional measures of performance on models used in AI-enabled autonomous systems can provide a false sense of confidence for systems operating in open-world contexts. Table~\ref{tab:cifar-accuracy} shows a set of nominal accuracy figures for a set of neural networks trained on the CIFAR-10 identification problem. However, these figures alone tell us little about the robustness of the models when deployed and therefore additional evidence that they are suitable for use in these complex contexts is required.
\vspace{3mm}
    \begin{center}
    \captionof{table}{CIFAR-10 model accuracy~\cite{paterson2021deepcert}\label{tab:cifar-accuracy}}
    \begin{footnotesize}
    \begin{tabular}{clcclcclc}
        \toprule
         \multicolumn{2}{c}{Model} & Accuracy& \multicolumn{2}{c}{Model} & Accuracy&  \multicolumn{2}{c}{Model} & Accuracy\\ \midrule
         4A & \multirow{2}{*}{Small Relu}  & 49.11&
         5A & \multirow{2}{*}{Large Relu}& 53.20&
         6A & \multirow{2}{*}{CNN} & 84.07\\
         4B &  & 47.45 &
         5B && 53.04&
         6B &  & 85.17\\ 
         \bottomrule
    \end{tabular}
    \end{footnotesize}
    \end{center}
    \vspace{3mm}
Verification is an important step in providing such evidence. Since vision-based systems have been shown to be susceptible to small perturbations in the input space~\cite{goodfellow2014explaining}, such an approach may be considered a measure of robustness. However, it lacks semantic meaning. What we need are tools and techniques which are of practical value to ML engineers, allowing them to build mitigations and operational safeguards which are aware of the limitations in robustness of the models used. Figure \ref{fig:ArchPattern} shows a process by which we may verify contextually meaningful measures of robustness~\cite{paterson2021deepcert}.

\vspace{3mm}
\begin{center}
    \includegraphics[width=0.6\textwidth]{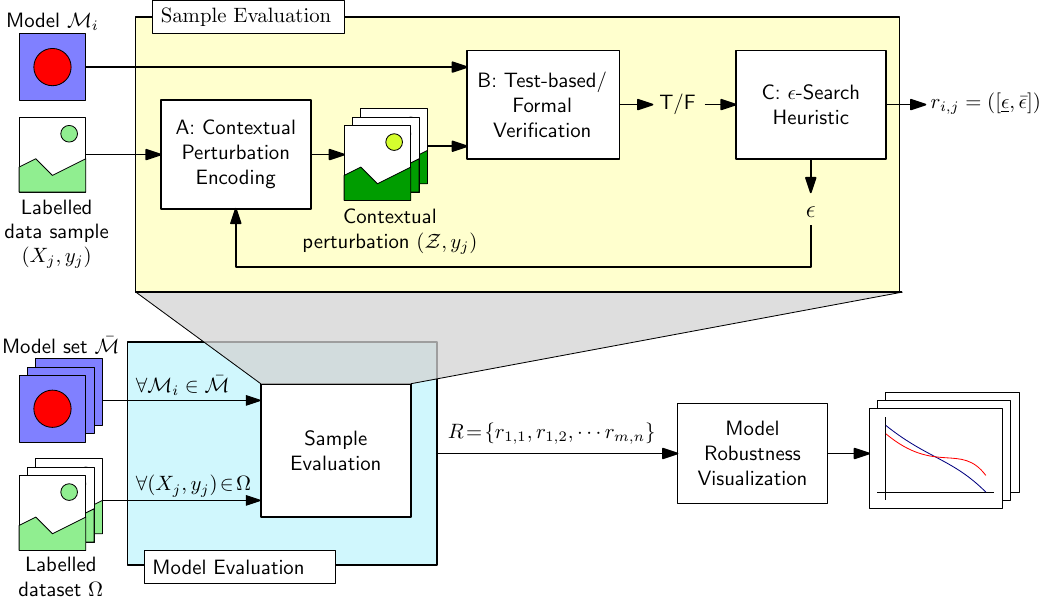}
\captionof{figure}{\label{fig:ArchPattern}Process for verifying contextually meaningful DNN robustness~\cite{paterson2021deepcert}}
\end{center}

In this work a set of contextually meaningful perturbations are identified, through discussions with domain experts, and a formal encoding of the perturbation defined such that $\epsilon \in [0, 1]$. Data samples are then perturbed with values of $\epsilon$ to gradually degrade the samples. We can then identify the level of perturbation necessary for a model to fail for a sample.

Figure~\ref{fig:deepcert} shows the result of this process. As the levels of contextually meaningful perturbation  (haze, contrast and blur) present in the image increase we see a corresponding degradation in the model performance. We note however that the rate of degradation is not constant across all models, indeed for haze, the ‘best model’ changes as the image degrades. Such evidence may lead us to refine our model in the presence of such conditions or to deploy multiple models with appropriate monitoring and switching to mitigate safety concerns.

\begin{minipage}{0.3\textwidth}
\centering
         \includegraphics[width=\textwidth]{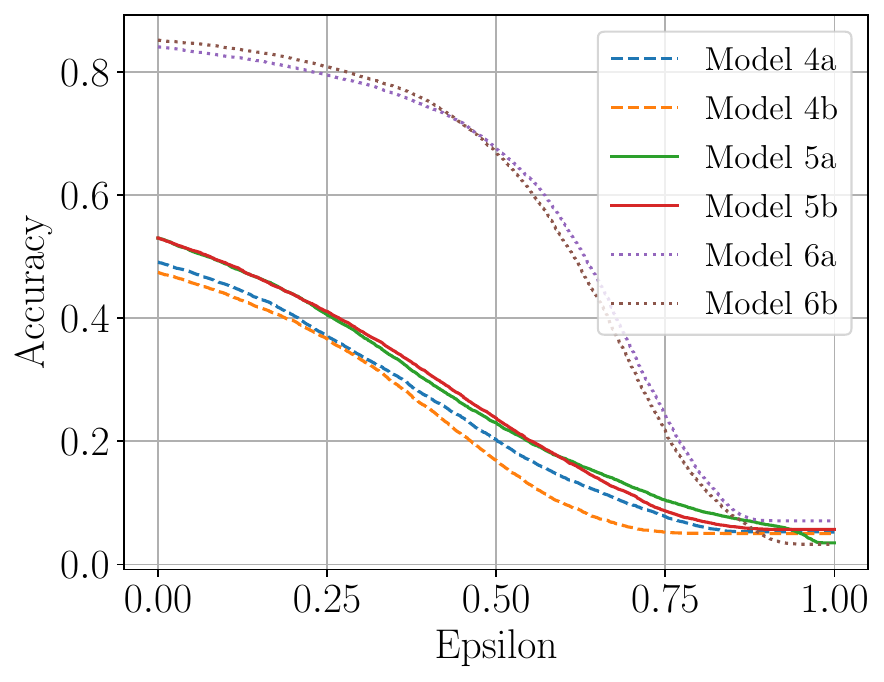}
Haze
\end{minipage}
\begin{minipage}{0.3\textwidth}
\centering
         \includegraphics[width=\textwidth]{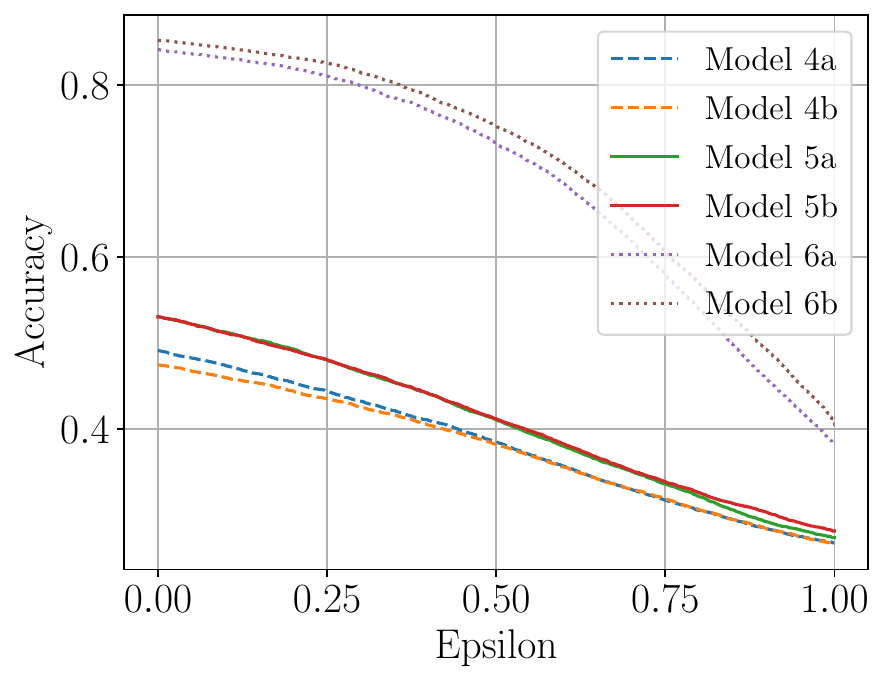}
Contrast
\end{minipage}
\begin{minipage}{0.3\textwidth}
\centering
         \includegraphics[width=\textwidth]{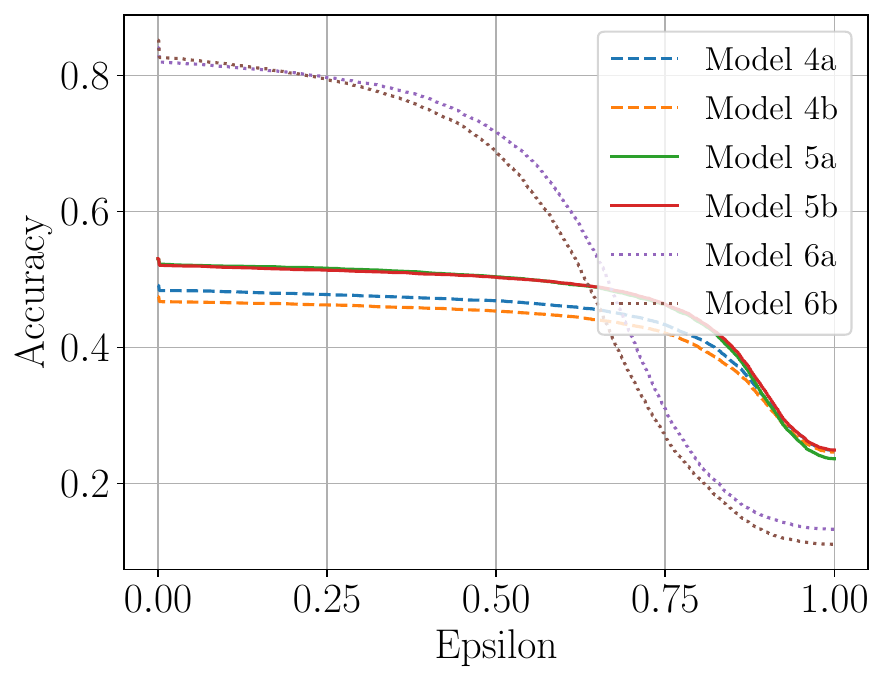}
Blur
\end{minipage}
         \captionof{figure}{Plot of model robustness for a set of CIFAR-10 models~\cite{paterson2021deepcert}\label{fig:deepcert}}

\end{example}

\section{AI Safety: The General-Purpose AI Model Safety Argument}
At this level of abstraction, the BIG argument considers the rapid adoption of foundational or frontier AI models, such as LLMs, including in critical applications such as healthcare \cite{gallifant2025tripod}\cite{thirunavukarasu2023large}. These models are often presented as general purpose, with the intended context rarely specified \cite{bengio2025international}. This makes their safety assurance at odds with long-established safety principles and practices that consider safety as a context-sensitive property \cite{habli2023meaning} \cite{mcdermid2024upstream}.

The general-purpose nature of foundational models has shifted the technical AI safety debate from context to capability \cite{amodei2016concrete}\cite{anderljung2023frontier}\cite{AISICapability25}. Major initiatives for assuring the safety of General-Purpose AI (GPAI) models have concentrated on the potential ‘\textit{harmful}’  outcomes that model capabilities may cause. For example, Google DeepMind’s Frontier Safety Framework focuses on different critical capability levels (CCLs) \cite{GoogleSafe2025}. CCLs describe “\textit{protocols for the detection of capability levels at which models may pose severe risks}” \cite{GoogleSafe2025}. 

The latest version (2.0) focuses almost exclusively on misuse and deceptive alignment risks. For example, for the former risk, the framework pays a particular attention to frontier models “\textit{assisting in the development, preparation, and/or execution of a chemical, biological, radiological, or nuclear (“CBRN”) attack}” \cite{GoogleSafe2025}. It considers the use of safety cases for assuring the sufficient mitigation of this risk at the development, pre-deployment and post-deployment stages.

The risk associated with these frontier model capabilities may be viewed as a common cause failure or a particular risk \cite{mcdermid2024upstream}. That is, these failure conditions are problematic regardless of the specific context or application (i.e. downstream safety). This has led to an emphasis in the technical AI safety literature on evaluation, independent audits and red-team testing conducted at the model development stage and at the capability level (i.e. upstream safety), where it is believed to be more effective to identify early warning signs (see Example \ref{exp:Anthropc} for an independent pre-deployment evaluation of Anthropic’s Claude 3.5 Sonnet \cite{AISIUSUK}).

\begin{figure}
  \centering
  \includegraphics[width=1\textwidth]{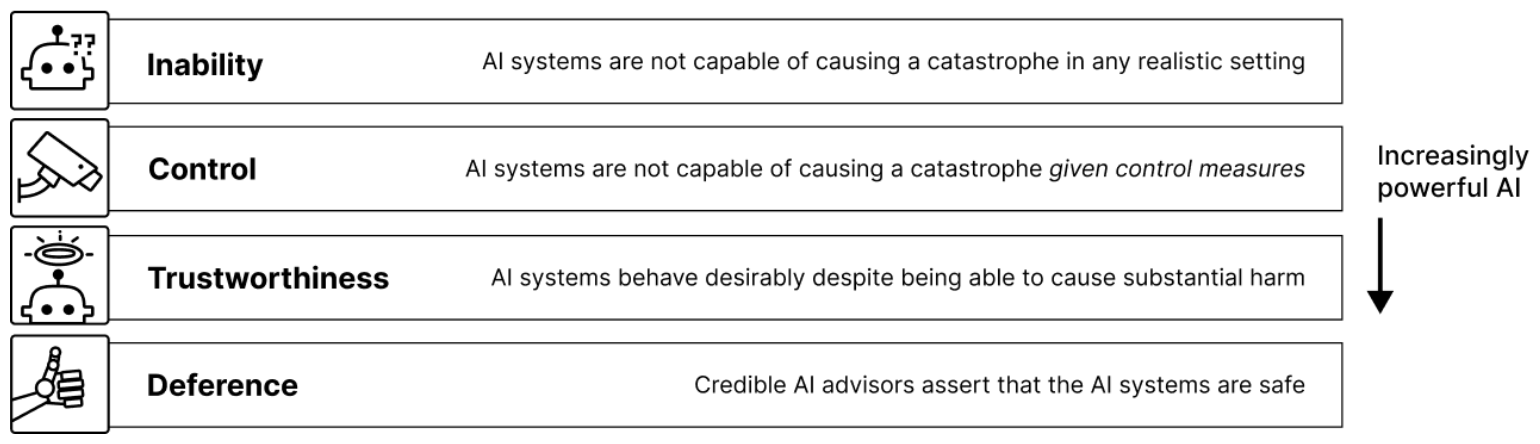}
  \caption{\label{fig:SCAdvAI} Building block arguments for making frontier AI safety cases \cite{clymer2024safety}}
\end{figure}

Figure \ref{fig:SCAdvAI} depicts a preliminary proposal developed by Clymer et al. \cite{clymer2024safety} for structuring a safety case for ‘\textit{advanced}’ AI systems. The ‘\textit{blocks}’ in this argument structure are based on a scale of dangerous capability. This ranges from assertions about the inability of the AI model to cause catastrophic events to the capability being controlled, trusted or monitored by a ‘\textit{credible}’ AI advisor. It is important to note that research into the development of safety arguments and patterns for frontier or foundational models remains in its early stages and is yet to be subjected to independent scrutiny.

\begin{figure}
  \centering
  \includegraphics[width=0.7\textwidth]{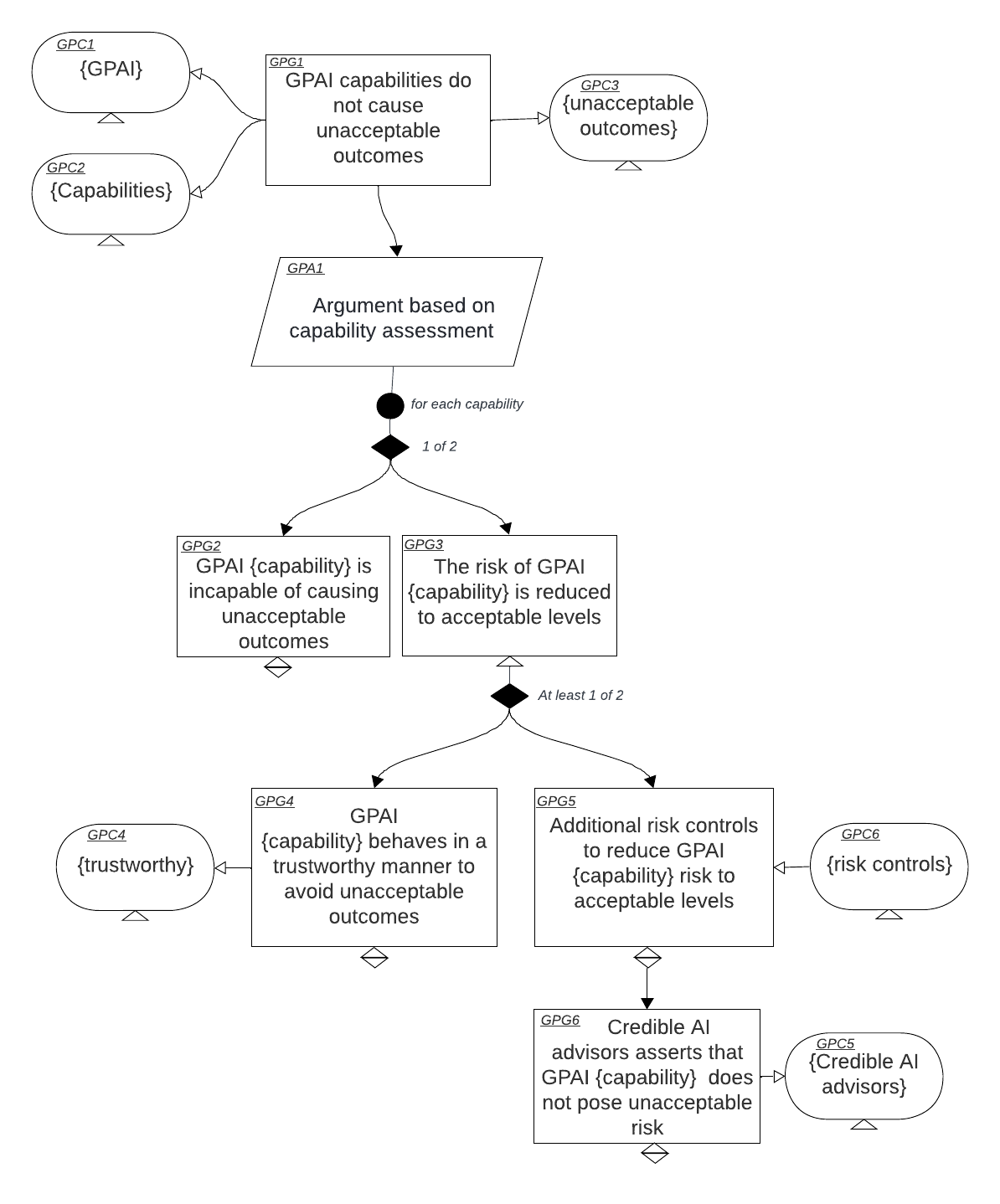}
  \caption{\label{fig:GPAIArg}A General-Purpose AI Model Safety Argument represented in GSN}
\end{figure}

In Figure \ref{fig:GPAIArg}, we adapted and remodeled the overarching capability-based safety argument (depicted in Figure \ref{fig:SCAdvAI}) as a GSN pattern. Essentially, the line of reasoning captured in the pattern is as follows: The top-level claim that “\textit{GPAI capabilities do not cause unacceptable outcomes}” (\textit{GPG1}), is supported by considering each capability, by arguing either that the capability is unable to cause unacceptable outcomes or that the risk of such outcomes is reduced via one or more controls, including through trustworthy behaviour (safe-by-design) or oversight by credible AI advisors (external controls). The expressiveness of the argument is improved by insisting on defining key concepts such as unacceptable outcomes (\textit{GPC3}) and credible AI advisors (\textit{GPC5}). 

A recent report by the UK AI Security Institute highlighted that developing a robust safety argument for frontier AI remains a significant challenge \cite{AISIUK2025}. This relates to (A) implementation issues such as the readiness of organisations to adopt safety case practices and integrate them into existing governance frameworks, (B) technical matters such as eliciting the capabilities of evolving AI models, and (C) the act of writing frontier AI safety cases, including deciding on the top-level claims and assessing confidence in the arguments and evidence \cite{barrett2025assessing}.

Assuringly, frontier AI companies have started to publish some of their initial or preliminary safety cases, most notably Anthropic \cite{AnthropicSketches}. Anthropic uses safety cases as a primary means for supporting the implementation of its ‘Responsible Scaling Policy’ (RSP) \cite{AnthropicRSP}. RSP represents the organisation's framework for managing risks from increasingly capable AI systems, defining risk thresholds after which model capabilities require safeguards to mitigate the risks. It is noteworthy that RSP advocates for proportional safeguards, i.e. "\textit{safeguards that scale with potential risks}" \cite{AnthropicRSP}. This is consistent with the BIG argument, and the approach adopted in traditional safety engineering for many years, within which the issue of risk mitigation and acceptance is not merely technical and requires an in-depth and inclusive consideration, trade-offs and justification from technical and sociotechnical perspectives \cite{lowrance1976acceptable}. This reinforces the need for integration and traceability between the different kinds of arguments within an AI safety case.

\begin{example} [Pre-Deployment Evaluation of Anthropic’s Claude 3.5 Sonnet]
\label{exp:Anthropc}
For a safety case to be complete, evidence must be provided to support the safety argument presented. For LLMs, evaluation via red-teaming is often presented as a key measure for generating the necessary evidence. Here, the example is based on “Pre-Deployment Evaluation of Anthropic’s Claude 3.5 Sonnet” \cite{AISIUSUK}. The evaluation was jointly conducted and reported by the U.K. AI Safety (now Security) Institute and U.S. AI Safety Institute. The evaluation considered different types of capabilities namely (1) biological capabilities, (2) cyber capabilities, (3) software and AI development and (4) safeguard efficacy. 

Considering the last category, Figure \ref{fig:Anthropic} shows the results reported by the U.S. AI Safety Institute, revealing that when subjected to ‘jailbreak’ attacks, the model can assist with requests that could potentially lead to harmful effects (based on different HarmBench categories) \cite{mazeika2024harmbench}. That is, the model may be vulnerable to jailbreaks despite the technical safeguards designed by the developers. It is important to note that technical safeguards in LLMs are just one of several risk control measures needed at different technology, system and societal levels. As our main argument shows, the sufficiency of these measures must be clearly justified, challenged and reviewed by the relevant stakeholders.

\vspace{3mm}
\begin{center}
    \includegraphics[width=0.8\textwidth]{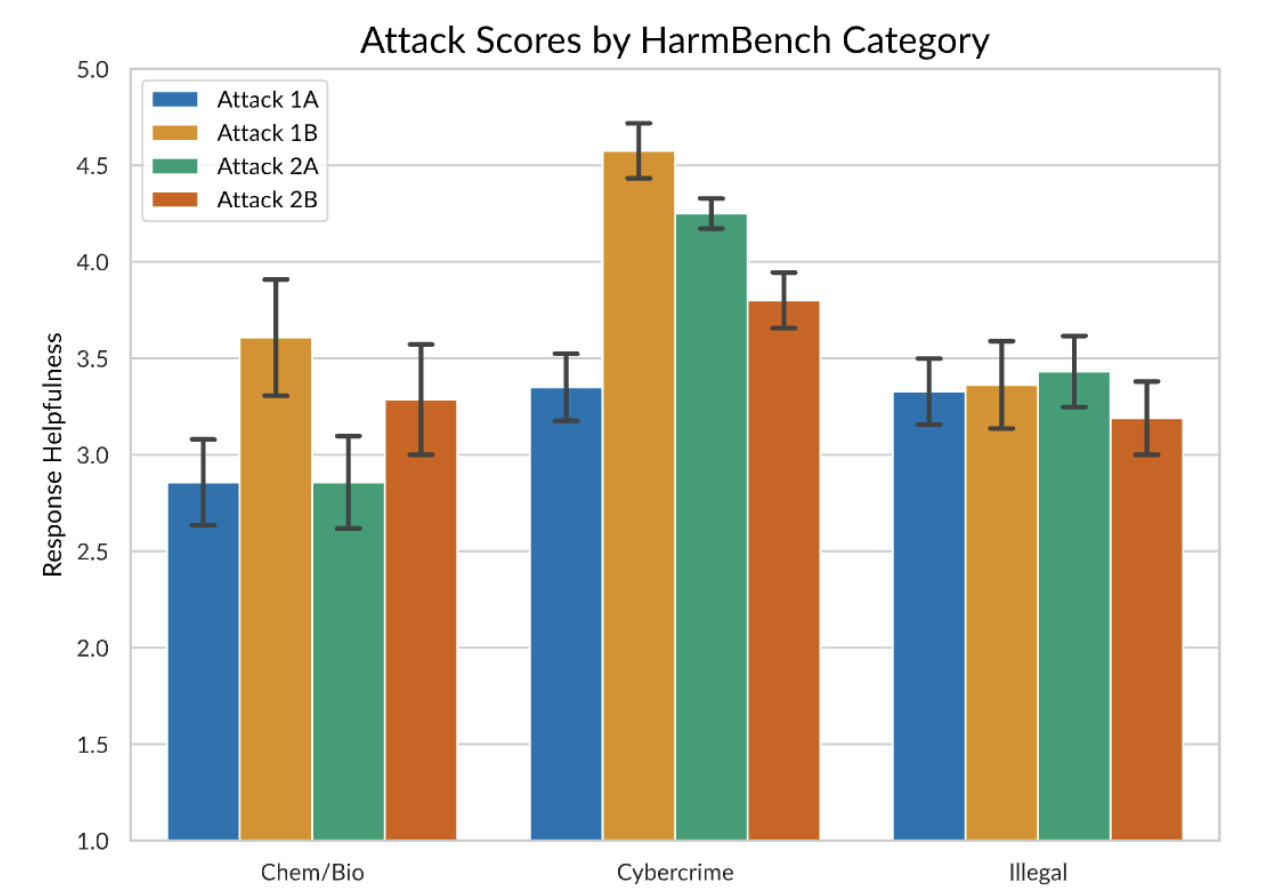}
\captionof{figure}{\label{fig:Anthropic}Anthropic’s Claude 3.5 Sonnet Performance against HarmBench Categories) \cite{AISIUSUK}}
\end{center}

\end{example}

\section{Concluding remarks}
The BIG argument reveals the complexity of the chain of reasoning and the scale of evidence necessary for assuring the safe deployment of AI systems in critical applications, including those utilising frontier models. The argument centers on three characteristics. Firstly, it is \textit{balanced} by addressing safety alongside other critical ethical issues such as privacy and equity, acknowledging complexities and trade-offs in the broader societal impact of AI. Secondly, it is \textit{integrated} by bringing together the social, ethical and technical aspects of safety assurance in a way that is traceable and accountable. Third, it is \textit{grounded} by building on long-established safety norms and practices from safety-critical systems, such as being sensitive to context and maintaining risk proportionality.

The BIG argument highlights the multidisciplinary, participatory and sociotechnical nature of safety assurance for complex AI-based systems, especially when granted more autonomy and deployed in open environments. We conclude with the following remarks:

\begin{itemize}
    \item \textbf{Beyond Many Extremes:} The BIG argument brings together different, complementary perspectives, avoiding unnecessary exceptionalism in the AI safety debate, such as technical vs. non-technical risks or catastrophic vs. systematic harms. While safety has historically focused on accidental harm, the cyber capabilities enabled by frontier AI and the security risks they pose reinforces the need for closer integration between AI safety and security assurance, possibly under the broader umbrella of resilience engineering.
    \item \textbf{Context is Key, but Capability Assurance is Essential:} Safety is context-sensitive. Effective safety risk assessment requires a sufficient characterisation of the intended environment. The BIG argument refines the notion of context across social, ethical, system and technological levels. However, frontier AI models produce general-purpose capabilities that are concerning regardless of the specific context, such as hiding behaviours during testing or undermining oversight. Instead of labelling these behaviours as safe or unsafe, we should ensure the models do not exhibit them, with high confidence, and assess safety risk as soon as the deployment context is determined and scoped.
    \item \textbf{Traceability for Accountability:} A core aspect of the BIG argument is traceability, maintaining a chain of reasoning that links the risk of harm to safety requirements and metrics driving the design and evaluation of AI models and their training and testing datasets. Ensuring traceability represents sound engineering and provides a basis for accountability throughout the AI lifecycle \cite{porter2022distinguishing}\cite{raji2020closing}\cite{cobbe2023understanding}.
    \item \textbf{Fast Rate of Change and Dynamism: }Given the rapid nature of AI development, it is important to integrate the BIG argument into a phased and iterative process that includes proactive monitoring and updates, ensuring the argument remains valid within a dynamic safety case that evolves with system and context changes \cite{denney2015dynamic}\cite{carlan2024dynamic}.
    \item \textbf{Urgent Need for Case Studies and Exemplars:} In the face of novelty, we seek comfort in first principles. However, this should be combined with case studies on the use of safety cases for actual AI systems in diverse domains and applications, contributing to a body of credible, peer-reviewed knowledge in safety cases and guidelines valuable to developers, users and policymakers \cite{AISIUK2025}.
\end{itemize}

We see the BIG argument as a step towards unifying the wide range of concerns related to the safe use of AI, especially frontier models. We also believe it will help shape the research agenda for AI Safety. In particular, we stress the importance of traceability for accountability. In traditional safety engineering, emphasis is placed on designing for safety, which is known to be both effective and cost-effective. Given the way frontier AI models are developed, this is perhaps one of the hardest objectives to achieve. The resolution might be to build on the upstream-downstream concepts \cite{mcdermid2024upstream}, with "\textit{design for safety}" shaping the use of frontier AI in its downstream context. This could be one of the most fruitful areas to develop use cases or exemplars.

\section{Acknowledgments}

This work was supported by the Centre for Assuring Autonomy, a partnership between Lloyd’s Register Foundation and the University of York, and the UKRI AI Centre for Doctoral Training in Safe Artificial Intelligence Systems (SAINTS) (EP/Y030540/1). Special thanks to Ana MacIntosh, Rob Alexander, Shaun Feakins and Mark Nicholson for their valuable feedback.

\bibliographystyle{unsrt}

\bibliography{sample-base.bib}

\begin{thebibliography}{100}

\bibitem{UNAI}
United Nations.
\newblock Governing {AI} for humanity: Final report.
\newblock \url{https://www.un.org/sites/un2.un.org/files/governing_ai_for_humanity_final_report_en.pdf}, 2024.
\newblock Accessed: 20 Feb 2025.

\bibitem{topol2019high}
Eric~J Topol.
\newblock High-performance medicine: the convergence of human and artificial intelligence.
\newblock {\em Nature medicine}, 25(1):44--56, 2019.

\bibitem{OECD1}
Eric~Sutherland Derya~Şahin.
\newblock With mounting pressure on health systems, can ai help 8 billion people to obtain optimal health outcomes?
\newblock \url{https://oecd.ai/en/wonk/optimal-health-outcomes}, March 2024.

\bibitem{zeng2024ai}
Yi~Zeng, Kevin Klyman, Andy Zhou, Yu~Yang, Minzhou Pan, Ruoxi Jia, Dawn Song, Percy Liang, and Bo~Li.
\newblock Ai risk categorization decoded (air 2024): From government regulations to corporate policies.
\newblock {\em arXiv preprint arXiv:2406.17864}, 2024.

\bibitem{lee2025saif}
Kyeongryul Lee, Heehyeon Kim, and Joyce~Jiyoung Whang.
\newblock Saif: A comprehensive framework for evaluating the risks of generative ai in the public sector.
\newblock {\em arXiv preprint arXiv:2501.08814}, 2025.

\bibitem{habli2024ai}
Ibrahim Habli and John~Alexander McDermid.
\newblock {AI} safety: Navigating the expanding landscape of potential harms.
\newblock {\em Safety-Critical Systems Club Newsletter. \url{https://scsc.uk/r194.5:1#page=29}}, 2024.

\bibitem{uuk2024taxonomysystemicrisksgeneralpurpose}
Risto Uuk, Carlos~Ignacio Gutierrez, Daniel Guppy, Lode Lauwaert, Atoosa Kasirzadeh, Lucia Velasco, Peter Slattery, and Carina Prunkl.
\newblock {A Taxonomy of Systemic Risks from General-Purpose AI}.
\newblock {\em arXiv preprint arXiv:2412.07780}, 2024.

\bibitem{chandra2024livedexperienceinsightunpacking}
Mohit Chandra, Suchismita Naik, Denae Ford, Ebele Okoli, Munmun De~Choudhury, Mahsa Ershadi, Gonzalo Ramos, Javier Hernandez, Ananya Bhattacharjee, Shahed Warreth, et~al.
\newblock From lived experience to insight: Unpacking the psychological risks of using ai conversational agents.
\newblock {\em arXiv preprint arXiv:2412.07951}, 2024.

\bibitem{stilgoe2021can}
Jack Stilgoe.
\newblock How can we know a self-driving car is safe?
\newblock {\em Ethics and Information Technology}, 23(4):635--647, 2021.

\bibitem{uber}
{National Transportation Safety Board}.
\newblock {Collision Between Vehicle Controlled by Developmental Automated Driving System and Pedestrian, Tempe, Arizona, March 18, 2018, NTSB/HAR-19/03}, 2019.

\bibitem{koopman2024anatomy}
Philip Koopman.
\newblock Anatomy of a robotaxi crash: Lessons from the cruise pedestrian dragging mishap.
\newblock In {\em International Conference on Computer Safety, Reliability, and Security}, pages 119--133. Springer, 2024.

\bibitem{vaswani2017attention}
A~Vaswani.
\newblock Attention is all you need.
\newblock {\em Advances in Neural Information Processing Systems}, 2017.

\bibitem{steyvers2025large}
Mark Steyvers, Heliodoro Tejeda, Aakriti Kumar, Catarina Belem, Sheer Karny, Xinyue Hu, Lukas~W Mayer, and Padhraic Smyth.
\newblock What large language models know and what people think they know.
\newblock {\em Nature Machine Intelligence}, pages 1--11, 2025.

\bibitem{wu2024inference}
Yangzhen Wu, Zhiqing Sun, Shanda Li, Sean Welleck, and Yiming Yang.
\newblock Inference scaling laws: An empirical analysis of compute-optimal inference for problem-solving with language models.
\newblock {\em arXiv preprint arXiv:2408.00724}, 2024.

\bibitem{bengio2025international}
Yoshua. Bengio~et al.
\newblock {International AI Safety Report}.
\newblock \url{https://www.gov.uk/government/publications/international-ai-safety-report-2025}, 2025.
\newblock DSIT 2025/001.

\bibitem{gibney2025china}
Elizabeth Gibney.
\newblock China’s cheap, open ai model deepseek thrills scientists.
\newblock {\em Nature}, 638(8049):13--14, 2025.

\bibitem{xie2025vlms}
Shaoyuan Xie, Lingdong Kong, Yuhao Dong, Chonghao Sima, Wenwei Zhang, Qi~Alfred Chen, Ziwei Liu, and Liang Pan.
\newblock Are vlms ready for autonomous driving? an empirical study from the reliability, data, and metric perspectives.
\newblock {\em arXiv preprint arXiv:2501.04003}, 2025.

\bibitem{yang2023llm4drive}
Zhenjie Yang, Xiaosong Jia, Hongyang Li, and Junchi Yan.
\newblock Llm4drive: A survey of large language models for autonomous driving.
\newblock In {\em NeurIPS 2024 Workshop on Open-World Agents}, 2023.

\bibitem{farquhar2024detecting}
Sebastian Farquhar, Jannik Kossen, Lorenz Kuhn, and Yarin Gal.
\newblock Detecting hallucinations in large language models using semantic entropy.
\newblock {\em Nature}, 630(8017):625--630, 2024.

\bibitem{jones2025ai}
Nicola Jones.
\newblock Ai hallucinations can't be stopped—but these techniques can limit their damage.
\newblock {\em Nature}, 637(8047):778--780, 2025.

\bibitem{fearnleyrisk}
Laura Fearnley, Elly Cairns, Tom Stoneham, Philippa Ryan, Jenn Chubb, Jo~Iacovides, Cynthia~Iglesias Urrutia, Phillip Morgan, John McDermid, and Ibrahim Habli.
\newblock Risk of what? defining harm in the context of ai safety.
\newblock \url{https://philarchive.org/archive/FEAROW}, 2025.

\bibitem{hendrycks2023overviewcatastrophicairisks}
Dan Hendrycks, Mantas Mazeika, and Thomas Woodside.
\newblock {An Overview of Catastrophic AI Risks}.
\newblock \url{https://arxiv.org/abs/2306.12001}, 2023.

\bibitem{10.1145/3600211.3604673}
Renee Shelby, Shalaleh Rismani, Kathryn Henne, AJung Moon, Negar Rostamzadeh, Paul Nicholas, N'Mah Yilla-Akbari, Jess Gallegos, Andrew Smart, Emilio Garcia, and Gurleen Virk.
\newblock Sociotechnical harms of algorithmic systems: Scoping a taxonomy for harm reduction.
\newblock In {\em Proceedings of the 2023 AAAI/ACM Conference on AI, Ethics, and Society}, AIES '23, page 723–741, New York, NY, USA, 2023. Association for Computing Machinery.

\bibitem{burton2020mind}
Simon Burton, Ibrahim Habli, Tom Lawton, John McDermid, Phillip Morgan, and Zoe Porter.
\newblock Mind the gaps: Assuring the safety of autonomous systems from an engineering, ethical, and legal perspective.
\newblock {\em Artificial Intelligence}, 279:103201, 2020.

\bibitem{chan2023harms}
Alan Chan, Rebecca Salganik, Alva Markelius, Chris Pang, Nitarshan Rajkumar, Dmitrii Krasheninnikov, Lauro Langosco, Zhonghao He, Yawen Duan, Micah Carroll, et~al.
\newblock Harms from increasingly agentic algorithmic systems.
\newblock In {\em Proceedings of the 2023 ACM Conference on Fairness, Accountability, and Transparency}, pages 651--666, 2023.

\bibitem{bengio2023managing}
Yoshua Bengio, Geoffrey Hinton, Andrew Yao, Dawn Song, Pieter Abbeel, Yuval~Noah Harari, Ya-Qin Zhang, Lan Xue, Shai Shalev-Shwartz, Gillian Hadfield, et~al.
\newblock Managing {AI} risks in an era of rapid progress.
\newblock {\em arXiv preprint arXiv:2310.17688}, 2023.

\bibitem{lazar2023ai}
Seth Lazar and Alondra Nelson.
\newblock Ai safety on whose terms?
\newblock {\em Science}, 381(6654):138--138, 2023.

\bibitem{habli2020artificial}
Ibrahim Habli, Tom Lawton, and Zoe Porter.
\newblock Artificial intelligence in health care: accountability and safety.
\newblock {\em Bulletin of the World Health Organization}, 98(4):251, 2020.

\bibitem{koopman2024redefining}
Philip Koopman and William Widen.
\newblock Redefining safety for autonomous vehicles.
\newblock In {\em International Conference on Computer Safety, Reliability, and Security}, pages 300--314. Springer, 2024.

\bibitem{porter2018moral}
Zo{\"e} Porter, Ibrahim Habli, Helen Monkhouse, and John Bragg.
\newblock The moral responsibility gap and the increasing autonomy of systems.
\newblock In {\em Computer Safety, Reliability, and Security: SAFECOMP 2018 Workshops, ASSURE, DECSoS, SASSUR, STRIVE, and WAISE, V{\"a}ster{\aa}s, Sweden, September 18, 2018, Proceedings 37}, pages 487--493. Springer, 2018.

\bibitem{bainbridge1983ironies}
Lisanne Bainbridge.
\newblock Ironies of automation.
\newblock In {\em Analysis, design and evaluation of man--machine systems}, pages 129--135. Elsevier, 1983.

\bibitem{leveson2016engineering}
Nancy~G Leveson.
\newblock {\em Engineering a safer world: Systems thinking applied to safety}.
\newblock The MIT Press, 2016.

\bibitem{monkhouse2020enhanced}
Helen~E Monkhouse, Ibrahim Habli, and John McDermid.
\newblock An enhanced vehicle control model for assessing highly automated driving safety.
\newblock {\em Reliability Engineering \& System Safety}, 202:107061, 2020.

\bibitem{sujan2019human}
Mark Sujan, Dominic Furniss, Kath Grundy, Howard Grundy, David Nelson, Matthew Elliott, Sean White, Ibrahim Habli, and Nick Reynolds.
\newblock Human factors challenges for the safe use of artificial intelligence in patient care.
\newblock {\em BMJ health \& care informatics}, 26(1):e100081, 2019.

\bibitem{kelly1999arguing}
Timothy~Patrick Kelly et~al.
\newblock {\em Arguing safety: a systematic approach to managing safety cases}.
\newblock PhD thesis, Citeseer, 1999.

\bibitem{bishop2000methodology}
Peter Bishop and Robin Bloomfield.
\newblock A methodology for safety case development.
\newblock In {\em Safety and Reliability}, volume~20, pages 34--42. Taylor \& Francis, 2000.

\bibitem{sujan2016should}
Mark~A Sujan, Ibrahim Habli, Tim~P Kelly, Simone Pozzi, and Christopher~W Johnson.
\newblock Should healthcare providers do safety cases? lessons from a cross-industry review of safety case practices.
\newblock {\em Safety science}, 84:181--189, 2016.

\bibitem{m2017a}
UK~MOD.
\newblock Safety management requirements for defence systems part 1: Requirements.
\newblock Standard {Def Stan} 00-56:2017, UK Ministry of Defence, 2017.

\bibitem{sujan2024changing}
Mark Sujan and Ibrahim Habli.
\newblock Changing the patient safety mindset: can safety cases help?
\newblock {\em BMJ Quality \& Safety}, 33(3):145--148, 2024.

\bibitem{hawkins2021guidance}
Richard Hawkins, Colin Paterson, Chiara Picardi, Yan Jia, Radu Calinescu, and Ibrahim Habli.
\newblock Guidance on the assurance of machine learning in autonomous systems (amlas).
\newblock {\em arXiv preprint arXiv:2102.01564}, 2021.

\bibitem{buhl2024safety}
Marie~Davidsen Buhl, Gaurav Sett, Leonie Koessler, Jonas Schuett, and Markus Anderljung.
\newblock Safety cases for frontier ai.
\newblock {\em arXiv preprint arXiv:2410.21572}, 2024.

\bibitem{burton2017making}
Simon Burton, Lydia Gauerhof, and Christian Heinzemann.
\newblock Making the case for safety of machine learning in highly automated driving.
\newblock In {\em Computer Safety, Reliability, and Security: SAFECOMP 2017 Workshops, ASSURE, DECSoS, SASSUR, TELERISE, and TIPS, Trento, Italy, September 12, 2017, Proceedings 36}, pages 5--16. Springer, 2017.

\bibitem{clymer2024safety}
Joshua Clymer, Nick Gabrieli, David Krueger, and Thomas Larsen.
\newblock Safety cases: How to justify the safety of advanced ai systems.
\newblock {\em arXiv preprint arXiv:2403.10462}, 2024.

\bibitem{goemans2024safety}
Arthur Goemans, Marie~Davidsen Buhl, Jonas Schuett, Tomek Korbak, Jessica Wang, Benjamin Hilton, and Geoffrey Irving.
\newblock Safety case template for frontier ai: A cyber inability argument.
\newblock {\em arXiv preprint arXiv:2411.08088}, 2024.

\bibitem{moller2018handbook}
Niklas M{\"o}ller, Sven~Ove Hansson, Jan-Erik Holmberg, and Carl Rollenhagen.
\newblock {\em Handbook of safety principles}, volume~9.
\newblock John Wiley \& Sons, 2018.

\bibitem{kelly2004systematic}
Tim Kelly.
\newblock A systematic approach to safety case management.
\newblock {\em SAE transactions}, pages 257--266, 2004.

\bibitem{porter2024principles}
Zoe Porter, Ibrahim Habli, John McDermid, and Marten Kaas.
\newblock A principles-based ethics assurance argument pattern for ai and autonomous systems.
\newblock {\em AI and Ethics}, 4(2):593--616, 2024.

\bibitem{hawkins2022guidance}
Richard Hawkins, Matt Osborne, Mike Parsons, Mark Nicholson, John McDermid, and Ibrahim Habli.
\newblock Guidance on the safety assurance of autonomous systems in complex environments (sace).
\newblock {\em arXiv preprint arXiv:2208.00853}, 2022.

\bibitem{korbak2025sketch}
Tomek Korbak, Joshua Clymer, Benjamin Hilton, Buck Shlegeris, and Geoffrey Irving.
\newblock A sketch of an ai control safety case.
\newblock {\em arXiv preprint arXiv:2501.17315}, 2025.

\bibitem{habli2021safety}
Ibrahim Habli, Rob Alexander, and Richard~David Hawkins.
\newblock Safety cases: An impending crisis?
\newblock In {\em Safety-Critical Systems Symposium (SSS’21)}, 2021.

\bibitem{arnold2016windscale}
Lorna Arnold.
\newblock {\em Windscale 1957: anatomy of a nuclear accident}.
\newblock Springer, 2016.

\bibitem{bishop2004future}
Peter Bishop, Robin Bloomfield, and Sofia Guerra.
\newblock The future of goal-based assurance cases.
\newblock In {\em Proc. Workshop on Assurance Cases}, pages 390--395, 2004.

\bibitem{kelly2005goal}
TP~Kelly, JA~McDermid, and RA~Weaver.
\newblock Goal-based safety standards: opportunities and challenges.
\newblock In {\em 23rd International System Safety Conference, San Diego, California}, 2005.

\bibitem{CullenPiper}
Lord~Douglas Cullen.
\newblock {\em The Public Inquiry into the Piper Alpha Disaster}.
\newblock London: HMSO, 1990.

\bibitem{Kingscross}
British~Transport Police.
\newblock The king's cross fire of 1987.
\newblock \url{https://www.btp.police.uk/police-forces/british-transport-police/areas/about-us/about-us/our-history/the-kings-cross-fire-of-1987/}.
\newblock Accessed: 27 Feb 2025.

\bibitem{Clapham}
Department of~Transport.
\newblock Investigation into the clapham junction railway accident.
\newblock \url{https://www.railwaysarchive.co.uk/documents/DoT_Hidden001.pdf/}.
\newblock Accessed: 27 Feb 2025.

\bibitem{Grenfell}
Martin Moore-Bick.
\newblock The grenfell tower inquiry.
\newblock \url{https://www.grenfelltowerinquiry.org.uk/}.
\newblock Accessed: 27 Feb 2025.

\bibitem{kelly2008safety}
Tim Kelly.
\newblock Are safety cases working.
\newblock {\em Safety Critical Systems Club Newsletter}, 17(2):31--33, 2008.

\bibitem{haddon2009nimrod}
Charles Haddon-Cave.
\newblock {\em The Nimrod Review: an independent review into the broader issues surrounding the loss of the RAF Nimrod MR2 aircraft XV230 in Afghanistan in 2006, report}.
\newblock London: The Stationery Office, 2009.

\bibitem{leveson2011use}
Nancy~G Leveson.
\newblock The use of safety cases in certification and regulation.
\newblock \url{https://dspace.mit.edu/handle/1721.1/102833}, 2011.
\newblock Massachusetts Institute of Technology. Engineering Systems Division.

\bibitem{graydon2020towards}
Mallory~Suzanne Graydon.
\newblock Towards efficacy hypotheses for safety cases.
\newblock In {\em 2020 16th European Dependable Computing Conference (EDCC)}, pages 51--58. IEEE, 2020.

\bibitem{rinehart2017understanding}
David~J Rinehart, John~C Knight, and Jonathan Rowanhill.
\newblock Understanding what it means for assurance cases to" work".
\newblock Technical report, 2017.

\bibitem{holloway2008safety}
Safety case notations: Alternatives for the non-graphically inclined?
\newblock In {\em 2008 3rd IET international conference on system safety}, pages 1--6. IET, 2008.

\bibitem{holloway2023friendly}
C~Michael Holloway.
\newblock The friendly argument notation (fan): 2023 version.
\newblock Technical report, National Aeronautics and Space Administration, 2023.

\bibitem{assurance2018goal}
The Assurance Case~Working GroupG.
\newblock Goal structuring notation community standard (version 2).
\newblock \url{https://scsc.uk/r141B:1?t=1}, 2018.

\bibitem{netkachova2015tool}
Kateryna Netkachova, Oleksandr Netkachov, and Robin Bloomfield.
\newblock Tool support for assurance case building blocks: Providing a helping hand with cae.
\newblock In {\em Computer Safety, Reliability, and Security: SAFECOMP 2015 Workshops, ASSURE, DECSoS. ISSE, ReSA4CI, and SASSUR, Delft, The Netherlands, September 22, 2015, Proceedings 34}, pages 62--71. Springer, 2015.

\bibitem{graydon2007assurance}
Patrick~J Graydon, John~C Knight, and Elisabeth~A Strunk.
\newblock Assurance based development of critical systems.
\newblock In {\em 37th Annual IEEE/IFIP International Conference on Dependable Systems and Networks (DSN'07)}, pages 347--357. IEEE, 2007.

\bibitem{wei2019model}
Ran Wei, Tim~P Kelly, Xiaotian Dai, Shuai Zhao, and Richard Hawkins.
\newblock Model based system assurance using the structured assurance case metamodel.
\newblock {\em Journal of Systems and Software}, 154:211--233, 2019.

\bibitem{denney2018tool}
Ewen Denney and Ganesh Pai.
\newblock Tool support for assurance case development.
\newblock {\em Automated Software Engineering}, 25(3):435--499, 2018.

\bibitem{carlan2022automating}
Carmen C{\^a}rlan, Lydia Gauerhof, Barbara Gallina, and Simon Burton.
\newblock Automating safety argument change impact analysis for machine learning components.
\newblock In {\em 2022 IEEE 27th Pacific Rim International Symposium on Dependable Computing (PRDC)}, pages 43--53. IEEE, 2022.

\bibitem{rushby2009formalism}
John Rushby.
\newblock Formalism in safety cases.
\newblock In {\em Making Systems Safer: Proceedings of the Eighteenth Safety-Critical Systems Symposium, Bristol, UK, 9-11th February 2010}, pages 3--17. Springer, 2009.

\bibitem{kelly2001concepts}
Tim Kelly.
\newblock Concepts and principles of compositional safety case construction.
\newblock {\em Contract Research Report for QinetiQ COMSA/2001/1/1}, 34, 2001.

\bibitem{fenn2007safety}
Jane Fenn, Richard Hawkins, Phil Williams, and Tim Kelly.
\newblock Safety case composition using contracts-refinements based on feedback from an industrial case study.
\newblock In {\em The Safety of Systems: Proceedings of the Fifteenth Safety-critical Systems Symposium, Bristol, UK, 13--15 February 2007}, pages 133--146. Springer, 2007.

\bibitem{denney2011towards}
Ewen Denney, Ganesh Pai, and Ibrahim Habli.
\newblock Towards measurement of confidence in safety cases.
\newblock In {\em 2011 International Symposium on Empirical Software Engineering and Measurement}, pages 380--383. IEEE, 2011.

\bibitem{hawkins2011new}
Richard Hawkins, Tim Kelly, John Knight, and Patrick Graydon.
\newblock A new approach to creating clear safety arguments.
\newblock In {\em Advances in Systems Safety: Proceedings of the Nineteenth Safety-Critical Systems Symposium, Southampton, UK, 8-10th February 2011}, pages 3--23. Springer, 2011.

\bibitem{goodenough2012toward}
John~B Goodenough, Charles~B Weinstock, and Ari~Z Klein.
\newblock Toward a theory of assurance case confidence.
\newblock {\em Pittsburgh, PA: Software Engineering Institute, Carnegie Mellon University}, 2012.

\bibitem{bloomfield2024confidence}
Robin Bloomfield and John Rushby.
\newblock Confidence in assurance 2.0 cases.
\newblock In {\em The Practice of Formal Methods: Essays in Honour of Cliff Jones, Part I}, pages 1--23. Springer, 2024.

\bibitem{guiochet2015model}
J{\'e}r{\'e}mie Guiochet, Quynh~Anh Do~Hoang, and Mohamed Kaaniche.
\newblock A model for safety case confidence assessment.
\newblock In {\em Computer Safety, Reliability, and Security: 34th International Conference, SAFECOMP 2015, Delft, The Netherlands, September 23-25, 2015, Proceedings 34}, pages 313--327. Springer, 2015.

\bibitem{kelly1997safety}
Tim~P Kelly and John~A McDermid.
\newblock Safety case construction and reuse using patterns.
\newblock In {\em Safe Comp 97: The 16th International Conference on Computer Safety, Reliability and Security}, pages 55--69. Springer, 1997.

\bibitem{ye2005justifying}
Fan Ye.
\newblock {\em Justifying the use of COTS Components within safety critical applications}.
\newblock PhD thesis, Citeseer, 2005.

\bibitem{denney2015dynamic}
Ewen Denney, Ganesh Pai, and Ibrahim Habli.
\newblock Dynamic safety cases for through-life safety assurance.
\newblock In {\em 2015 IEEE/ACM 37th IEEE International Conference on Software Engineering}, volume~2, pages 587--590. IEEE, 2015.

\bibitem{calinescu2017engineering}
Radu Calinescu, Danny Weyns, Simos Gerasimou, Muhammad~Usman Iftikhar, Ibrahim Habli, and Tim Kelly.
\newblock Engineering trustworthy self-adaptive software with dynamic assurance cases.
\newblock {\em IEEE Transactions on Software Engineering}, 44(11):1039--1069, 2017.

\bibitem{asaadi2020dynamic}
Erfan Asaadi, Ewen Denney, Jonathan Menzies, Ganesh~J Pai, and Dimo Petroff.
\newblock Dynamic assurance cases: a pathway to trusted autonomy.
\newblock {\em Computer}, 53(12):35--46, 2020.

\bibitem{codaf_safecomp}
Philippa Ryan, Sepeedeh Shahbeigi, Jie Zou, Ioannis Stefanakos, and John Molloy.
\newblock A dynamic assurance framework for an autonomous survey drone.
\newblock In Andrea Ceccarelli, Mario Trapp, Andrea Bondavalli, and Friedemann Bitsch, editors, {\em Computer Safety, Reliability, and Security}, pages 285--299, Cham, 2024. Springer Nature Switzerland.

\bibitem{rushby2015interpretation}
John Rushby.
\newblock The interpretation and evaluation of assurance cases.
\newblock {\em Comp. Science Laboratory, SRI International, Tech. Rep. SRI-CSL-15-01}, 2015.

\bibitem{holloway2015explicate}
C~Michael Holloway.
\newblock Explicate'78: Uncovering the implicit assurance case in do-178c.
\newblock In {\em Safety-Critical Systems Symposium 2015 (SSS 2015)}, number NF1676L-20463, 2015.

\bibitem{mohamad2021security}
Mazen Mohamad, Jan-Philipp Stegh{\"o}fer, and Riccardo Scandariato.
\newblock Security assurance cases—state of the art of an emerging approach.
\newblock {\em Empirical software engineering}, 26(4):70, 2021.

\bibitem{alexander2011security}
Rob Alexander, Richard Hawkins, and Tim Kelly.
\newblock Security assurance cases: motivation and the state of the art.
\newblock \url{https://www-users.york.ac.uk/~rdh2/papers/York%20CESG%20security%20case%20report.pdf}, 2011.
\newblock High Integrity Systems Engineering Department of Computer Science University of York Deramore Lane York YO10 5GH.

\bibitem{burr2023ethical}
Christopher Burr and David Leslie.
\newblock Ethical assurance: a practical approach to the responsible design, development, and deployment of data-driven technologies.
\newblock {\em AI and Ethics}, 3(1):73--98, 2023.

\bibitem{gauerhof2020assuring}
Lydia Gauerhof, Richard Hawkins, Chiara Picardi, Colin Paterson, Yuki Hagiwara, and Ibrahim Habli.
\newblock Assuring the safety of machine learning for pedestrian detection at crossings.
\newblock In {\em Computer Safety, Reliability, and Security: 39th International Conference, SAFECOMP 2020, Lisbon, Portugal, September 16--18, 2020, Proceedings 39}, pages 197--212. Springer, 2020.

\bibitem{lawton2024clinicians}
Tom Lawton, Phillip Morgan, Zoe Porter, Shireen Hickey, Alice Cunningham, Nathan Hughes, Ioanna Iacovides, Yan Jia, Vishal Sharma, and Ibrahim Habli.
\newblock Clinicians risk becoming ‘liability sinks’ for artificial intelligence.
\newblock {\em Future Healthcare Journal}, 11(1), 2024.

\bibitem{birch2022clinical}
Jonathan Birch, Kathleen~A Creel, Abhinav~K Jha, and Anya Plutynski.
\newblock Clinical decisions using ai must consider patient values.
\newblock {\em Nature {M}edicine}, 28(2):229--232, 2022.

\bibitem{burton2023addressing}
Simon Burton and Benjamin Herd.
\newblock Addressing uncertainty in the safety assurance of machine-learning.
\newblock {\em Frontiers in Computer Science}, 5:1132580, 2023.

\bibitem{koopman2023ul}
Philip Koopman.
\newblock Ul 4600: what to include in an autonomous vehicle safety case.
\newblock {\em Computer}, 56(05):101--104, 2023.

\bibitem{borg2023ergo}
Markus Borg, Jens Henriksson, Kasper Socha, Olof Lennartsson, Elias Sonnsj{\"o}~L{\"o}negren, Thanh Bui, Piotr Tomaszewski, Sankar~Raman Sathyamoorthy, Sebastian Brink, and Mahshid Helali~Moghadam.
\newblock Ergo, smirk is safe: a safety case for a machine learning component in a pedestrian automatic emergency brake system.
\newblock {\em Software quality journal}, 31(2):335--403, 2023.

\bibitem{jia2021safety}
Yan Jia, Tom Lawton, John Burden, John McDermid, and Ibrahim Habli.
\newblock Safety-driven design of machine learning for sepsis treatment.
\newblock {\em Journal of Biomedical Informatics}, 117:103762, 2021.

\bibitem{denney2023assurance}
Ewen Denney and Ganesh Pai.
\newblock Assurance-driven design of machine learning-based functionality in an aviation systems context.
\newblock In {\em 2023 IEEE/AIAA 42nd Digital Avionics Systems Conference (DASC)}, pages 1--10. IEEE, 2023.

\bibitem{fenn2023architecting}
Jane Fenn, Mark Nicholson, Ganesh Pai, and Michael Wilkinson.
\newblock Architecting safer autonomous aviation systems.
\newblock {\em arXiv preprint arXiv:2301.08138}, 2023.

\bibitem{mcdermid2024upstream}
John McDermid, Yan Jia, and Ibrahim Habli.
\newblock Upstream and downstream ai safety: Both on the same river?
\newblock {\em arXiv preprint arXiv:2501.05455}, 2024.

\bibitem{ISR-Safety}
DSIT Research~Paper Series.
\newblock International scientific report on the safety of advanced ai: interim report.
\newblock \url{https://assets.publishing.service.gov.uk/media/6716673b96def6d27a4c9b24/international_scientific_report_on_the_safety_of_advanced_ai_interim_report.pdf}, 2024.

\bibitem{hansson2018perform}
Sven~Ove Hansson.
\newblock How to perform an ethical risk analysis ({eRA}).
\newblock {\em Risk Analysis}, 38(9):1820--1829, 2018.

\bibitem{bender2021dangers}
Emily~M Bender, Timnit Gebru, Angelina McMillan-Major, and Shmargaret Shmitchell.
\newblock On the dangers of stochastic parrots: Can language models be too big?
\newblock In {\em Proceedings of the 2021 ACM conference on fairness, accountability, and transparency}, pages 610--623, 2021.

\bibitem{ganapathi2022tackling}
Shaswath Ganapathi, Jo~Palmer, Joseph~E Alderman, Melanie Calvert, Cyrus Espinoza, Jacqui Gath, Marzyeh Ghassemi, Katherine Heller, Francis Mckay, Alan Karthikesalingam, et~al.
\newblock Tackling bias in ai health datasets through the standing together initiative.
\newblock {\em Nature Medicine}, 28(11):2232--2233, 2022.

\bibitem{leslie2024ai}
David Leslie, Cami Rincon, Morgan Briggs, Antonella Perini, Smera Jayadeva, Ann Borda, SJ~Bennett, Christopher Burr, Mhairi Aitken, Michael Katell, et~al.
\newblock Ai fairness in practice.
\newblock {\em arXiv preprint arXiv:2403.14636}, 2024.

\bibitem{prunkl2024human}
Carina Prunkl.
\newblock Human autonomy at risk? an analysis of the challenges from ai.
\newblock {\em Minds and Machines}, 34(3):26, 2024.

\bibitem{dobbe2022system}
Roel Dobbe.
\newblock System safety and artificial intelligence.
\newblock In {\em Proceedings of the 2022 ACM Conference on Fairness, Accountability, and Transparency}, pages 1584--1584, 2022.

\bibitem{komorowski2018artificial}
Matthieu Komorowski, Leo~A Celi, Omar Badawi, Anthony~C Gordon, and A~Aldo Faisal.
\newblock The artificial intelligence clinician learns optimal treatment strategies for sepsis in intensive care.
\newblock {\em Nature medicine}, 24(11):1716--1720, 2018.

\bibitem{ashmore_assuring_2021}
Rob Ashmore, Radu Calinescu, and Colin Paterson.
\newblock Assuring the {Machine} {Learning} {Lifecycle}: {Desiderata}, {Methods}, and {Challenges}.
\newblock {\em ACM Computing Surveys}, 54(5):111:1--111:39, May 2021.

\bibitem{rasmussen1997risk}
Jens Rasmussen.
\newblock Risk management in a dynamic society: a modelling problem.
\newblock {\em Safety science}, 27(2-3):183--213, 1997.

\bibitem{leveson2023introduction}
Nancy~G Leveson.
\newblock {\em An introduction to system safety engineering}.
\newblock The MIT Press, 2023.

\bibitem{dekker2016just}
Sidney Dekker.
\newblock {\em Just culture: Balancing safety and accountability}.
\newblock crc Press, 2016.

\bibitem{dekker2019foundations}
Sidney Dekker.
\newblock {\em Foundations of safety science: A century of understanding accidents and disasters}.
\newblock Routledge, 2019.

\bibitem{rae2020manifesto}
Andrew Rae, David Provan, Hossam Aboelssaad, and Rob Alexander.
\newblock A manifesto for reality-based safety science.
\newblock {\em Safety science}, 126:104654, 2020.

\bibitem{habli2020enhancing}
Ibrahim Habli, Rob Alexander, Richard Hawkins, Mark Sujan, John McDermid, Chiara Picardi, and Tom Lawton.
\newblock Enhancing covid-19 decision making by creating an assurance case for epidemiological models.
\newblock {\em BMJ Health \& Care Informatics}, 27(3):e100165, 2020.

\bibitem{lowrance1976acceptable}
William~W Lowrance.
\newblock {\em Of acceptable risk: Science and the determination of safety.}
\newblock ERIC, 1976.

\bibitem{hansson2003ethical}
Sven~Ove Hansson.
\newblock Ethical criteria of risk acceptance.
\newblock {\em Erkenntnis}, 59(3):291--309, 2003.

\bibitem{sujan2017can}
Mark~A Sujan, Ibrahim Habli, Tim~P Kelly, Astrid G{\"u}hnemann, Simone Pozzi, and Christopher~W Johnson.
\newblock How can health care organisations make and justify decisions about risk reduction? lessons from a cross-industry review and a health care stakeholder consensus development process.
\newblock {\em Reliability Engineering \& System Safety}, 161:1--11, 2017.

\bibitem{GenAI-def}
Future of~Life~Institute.
\newblock {General Purpose AI and the AI Act}.
\newblock \url{https://artificialintelligenceact.eu/wp-content/uploads/2022/05/General-Purpose-AI-and-the-AI-Act.pdf}, February 2022.

\bibitem{wei2023jailbroken}
Alexander Wei, Nika Haghtalab, and Jacob Steinhardt.
\newblock Jailbroken: How does llm safety training fail?
\newblock {\em Advances in Neural Information Processing Systems}, 36:80079--80110, 2023.

\bibitem{perez2022red}
Ethan Perez, Saffron Huang, Francis Song, Trevor Cai, Roman Ring, John Aslanides, Amelia Glaese, Nat McAleese, and Geoffrey Irving.
\newblock Red teaming language models with language models.
\newblock {\em arXiv preprint arXiv:2202.03286}, 2022.

\bibitem{AnthropicSketches}
Anthropic.
\newblock Three sketches of asl-4 safety case components.
\newblock \url{https://alignment.anthropic.com/2024/safety-cases}.
\newblock Accessed: 8 March 2025.

\bibitem{balesni2024towards}
Mikita Balesni, Marius Hobbhahn, David Lindner, Alexander Meinke, Tomek Korbak, Joshua Clymer, Buck Shlegeris, J{\'e}r{\'e}my Scheurer, Charlotte Stix, Rusheb Shah, et~al.
\newblock Towards evaluations-based safety cases for ai scheming.
\newblock {\em arXiv preprint arXiv:2411.03336}, 2024.

\bibitem{AISIUK2025}
Benjamin Hilton, Marie~Davidsen Buhl, Tomek Korbak, and Geoffrey Irving.
\newblock Safety cases: A scalable approach to frontier ai safety.
\newblock {\em arXiv preprint arXiv:2503.04744}, 2025.

\bibitem{dobbe2024toward}
Roel Dobbe and Anouk Wolters.
\newblock Toward sociotechnical ai: Mapping vulnerabilities for machine learning in context.
\newblock {\em Minds and Machines}, 34(2):1--51, 2024.

\bibitem{weidinger2023sociotechnical}
Laura Weidinger, Maribeth Rauh, Nahema Marchal, Arianna Manzini, Lisa~Anne Hendricks, Juan Mateos-Garcia, Stevie Bergman, Jackie Kay, Conor Griffin, Ben Bariach, et~al.
\newblock Sociotechnical safety evaluation of generative ai systems.
\newblock {\em arXiv preprint arXiv:2310.11986}, 2023.

\bibitem{DeepMind-CCL-def}
Google~Deep Mind.
\newblock Frontier safety framework - version 1.
\newblock \url{https://storage.googleapis.com/deepmind-media/DeepMind.com/Blog/introducing-the-frontier-safety-framework/fsf-technical-report.pdf}, February 2024.

\bibitem{gao2023retrieval}
Yunfan Gao, Yun Xiong, Xinyu Gao, Kangxiang Jia, Jinliu Pan, Yuxi Bi, Yi~Dai, Jiawei Sun, and Haofen Wang.
\newblock Retrieval-augmented generation for large language models: A survey.
\newblock {\em arXiv preprint arXiv:2312.10997}, 2023.

\bibitem{chowdhury2025astrid}
Mohita Chowdhury, Yajie~Vera He, Aisling Higham, and Ernest Lim.
\newblock Astrid--an automated and scalable triad for the evaluation of rag-based clinical question answering systems.
\newblock {\em arXiv preprint arXiv:2501.08208}, 2025.

\bibitem{morley2020initial}
Jessica Morley, Luciano Floridi, Libby Kinsey, and Anat Elhalal.
\newblock From what to how: an initial review of publicly available ai ethics tools, methods and research to translate principles into practices.
\newblock {\em Science and engineering ethics}, 26(4):2141--2168, 2020.

\bibitem{beauchamp1994principles}
Tom~L Beauchamp and James~F Childress.
\newblock {\em Principles of biomedical ethics}.
\newblock Edicoes Loyola, 1994.

\bibitem{kazim2021interrelation}
Emre Kazim and Adriano Koshiyama.
\newblock The interrelation between data and ai ethics in the context of impact assessments.
\newblock {\em AI and Ethics}, 1:219--225, 2021.

\bibitem{jobin2019global}
Anna Jobin, Marcello Ienca, and Effy Vayena.
\newblock The global landscape of ai ethics guidelines.
\newblock {\em Nature machine intelligence}, 1(9):389--399, 2019.

\bibitem{floridi2018ai4people}
Luciano Floridi, Josh Cowls, Monica Beltrametti, Raja Chatila, Patrice Chazerand, Virginia Dignum, Christoph Luetge, Robert Madelin, Ugo Pagallo, Francesca Rossi, et~al.
\newblock Ai4people—an ethical framework for a good ai society: opportunities, risks, principles, and recommendations.
\newblock {\em Minds and machines}, 28:689--707, 2018.

\bibitem{von2021transparency}
Warren~J Von~Eschenbach.
\newblock Transparency and the black box problem: Why we do not trust ai.
\newblock {\em Philosophy \& Technology}, 34(4):1607--1622, 2021.

\bibitem{miller2019explanation}
Tim Miller.
\newblock Explanation in artificial intelligence: Insights from the social sciences.
\newblock {\em Artificial intelligence}, 267:1--38, 2019.

\bibitem{grice1975logic}
Herbert~Paul Grice.
\newblock Logic and conversation.
\newblock {\em Syntax and semantics}, 3:43--58, 1975.

\bibitem{kaas2024assuring}
Marten~HL Kaas and Ibrahim Habli.
\newblock Assuring ai safety: fallible knowledge and the gricean maxims.
\newblock {\em AI and Ethics}, pages 1--14, 2024.

\bibitem{cath2016reflective}
Yuri Cath.
\newblock Reflective equilibrium.
\newblock {\em The Oxford handbook of philosophical methodology}, 1, 2016.

\bibitem{rawls2017theory}
John Rawls.
\newblock A theory of justice.
\newblock In {\em Applied ethics}, pages 21--29. Routledge, 2017.

\bibitem{townsend2022pluralistic}
Beverley Townsend, Colin Paterson, TT~Arvind, Gabriel Nemirovsky, Radu Calinescu, Ana Cavalcanti, Ibrahim Habli, and Alan Thomas.
\newblock From pluralistic normative principles to autonomous-agent rules.
\newblock {\em Minds and Machines}, 32(4):683--715, 2022.

\bibitem{jevtic2018personalized}
Aleksandar Jevti{\'c}, Andr{\'e}s~Flores Valle, Guillem Aleny{\`a}, Greg Chance, Praminda Caleb-Solly, Sanja Dogramadzi, and Carme Torras.
\newblock Personalized robot assistant for support in dressing.
\newblock {\em IEEE transactions on cognitive and developmental systems}, 11(3):363--374, 2018.

\bibitem{MASS_VOJKOVIC2020333}
Goran Vojković and Melita Milenković.
\newblock Autonomous ships and legal authorities of the ship master.
\newblock {\em Case Studies on Transport Policy}, 8(2):333--340, 2020.

\bibitem{MASS_GOERLANDT2020104758}
Floris Goerlandt.
\newblock Maritime autonomous surface ships from a risk governance perspective: Interpretation and implications.
\newblock {\em Safety Science}, 128:104758, 2020.

\bibitem{ryan2024bridging}
Philippa Ryan, Mathias von Essen, Liam Shackley, and John McDermid.
\newblock Bridging the reality gap: Assurable simulations for an ml-based inspection drone flight controller.
\newblock In {\em International Conference on Computer Safety, Reliability, and Security}, pages 412--424. Springer, 2024.

\bibitem{machin2016smof}
Mathilde Machin, J{\'e}r{\'e}mie Guiochet, H{\'e}l{\`e}ne Waeselynck, Jean-Paul Blanquart, Matthieu Roy, and Lola Masson.
\newblock Smof: A safety monitoring framework for autonomous systems.
\newblock {\em IEEE Transactions on Systems, Man, and Cybernetics: Systems}, 48(5):702--715, 2016.

\bibitem{hollnagel2018safety}
Erik Hollnagel.
\newblock {\em Safety-I and safety-II: the past and future of safety management}.
\newblock CRC press, 2018.

\bibitem{hollnagel2017fram}
Erik Hollnagel.
\newblock {\em FRAM: the functional resonance analysis method: modelling complex socio-technical systems}.
\newblock Crc Press, 2017.

\bibitem{furniss2020using}
Dominic Furniss, David Nelson, Ibrahim Habli, Sean White, Matthew Elliott, Nick Reynolds, and Mark Sujan.
\newblock Using fram to explore sources of performance variability in intravenous infusion administration in icu: A non-normative approach to systems contradictions.
\newblock {\em Applied ergonomics}, 86:103113, 2020.

\bibitem{hawkins2023creating}
Richard Hawkins, Chiara Picardi, Lucy Donnell, and Murray Ireland.
\newblock Creating a safety assurance case for a machine learned satellite-based wildfire detection and alert system.
\newblock {\em Journal of Intelligent \& Robotic Systems}, 108(3):47, 2023.

\bibitem{barmpoutis2020review}
Panagiotis Barmpoutis, Periklis Papaioannou, Kosmas Dimitropoulos, and Nikos Grammalidis.
\newblock A review on early forest fire detection systems using optical remote sensing.
\newblock {\em Sensors}, 20(22):6442, 2020.

\bibitem{paterson2021detection}
Colin Paterson, Radu Calinescu, and Chiara Picardi.
\newblock Detection and mitigation of rare subclasses in deep neural network classifiers.
\newblock In {\em 2021 IEEE International Conference on Artificial Intelligence Testing (AITest)}, pages 9--16. IEEE, 2021.

\bibitem{paterson2021deepcert}
Colin Paterson, Haoze Wu, John Grese, Radu Calinescu, Corina~S P{\u{a}}s{\u{a}}reanu, and Clark Barrett.
\newblock Deepcert: Verification of contextually relevant robustness for neural network image classifiers.
\newblock In {\em Computer Safety, Reliability, and Security: 40th International Conference, SAFECOMP 2021, York, UK, September 8--10, 2021, Proceedings 40}, pages 3--17. Springer, 2021.

\bibitem{goodfellow2014explaining}
Ian~J Goodfellow, Jonathon Shlens, and Christian Szegedy.
\newblock Explaining and harnessing adversarial examples.
\newblock {\em arXiv preprint arXiv:1412.6572}, 2014.

\bibitem{gallifant2025tripod}
Jack Gallifant, Majid Afshar, Saleem Ameen, Yindalon Aphinyanaphongs, Shan Chen, Giovanni Cacciamani, Dina Demner-Fushman, Dmitriy Dligach, Roxana Daneshjou, Chrystinne Fernandes, et~al.
\newblock The tripod-llm reporting guideline for studies using large language models.
\newblock {\em Nature Medicine}, pages 1--10, 2025.

\bibitem{thirunavukarasu2023large}
Arun~James Thirunavukarasu, Darren Shu~Jeng Ting, Kabilan Elangovan, Laura Gutierrez, Ting~Fang Tan, and Daniel Shu~Wei Ting.
\newblock Large language models in medicine.
\newblock {\em Nature medicine}, 29(8):1930--1940, 2023.

\bibitem{habli2023meaning}
{Habli, Ibrahim}.
\newblock {On the Meaning of AI Safety}.
\newblock \url{https://eprints.whiterose.ac.uk/204545}, 2025.

\bibitem{amodei2016concrete}
Dario Amodei, Chris Olah, Jacob Steinhardt, Paul Christiano, John Schulman, and Dan Man{\'e}.
\newblock Concrete problems in ai safety.
\newblock {\em arXiv preprint arXiv:1606.06565}, 2016.

\bibitem{anderljung2023frontier}
Markus Anderljung, Joslyn Barnhart, Anton Korinek, Jade Leung, Cullen O'Keefe, Jess Whittlestone, Shahar Avin, Miles Brundage, Justin Bullock, Duncan Cass-Beggs, et~al.
\newblock Frontier ai regulation: Managing emerging risks to public safety.
\newblock {\em arXiv preprint arXiv:2307.03718}, 2023.

\bibitem{AISICapability25}
{AI Security Institute}.
\newblock {Principles for Evaluating Misuse Safeguards of Frontier AI Systems}, 2025.

\bibitem{GoogleSafe2025}
Google DeepMind.
\newblock Frontier safety framework version 2.0.
\newblock \url{https://storage.googleapis.com/deepmind-media/DeepMind.com/Blog/updating-thefrontier-safety-framework/Frontier%20Safety%20Framework%202.0%20(1).pdf}, 2025.
\newblock Accessed: 20 Feb 2025.

\bibitem{AISIUSUK}
{AISI}.
\newblock {US AISI and UK AISI Joint Pre-Deployment Test Anthropic’s Claude 3.5 Sonnet (October 2024 Release)}.
\newblock \url{https://eprints.whiterose.ac.uk/204545/}, 2024.

\bibitem{barrett2025assessing}
Stephen Barrett, Philip Fox, Joshua Krook, Tuneer Mondal, Simon Mylius, and Alejandro Tlaie.
\newblock Assessing confidence in frontier ai safety cases.
\newblock {\em arXiv preprint arXiv:2502.05791}, 2025.

\bibitem{AnthropicRSP}
Anthropic.
\newblock Responsible scaling policy.
\newblock \url{https://www.anthropic.com/news/anthropics-responsible-scaling-policy}.
\newblock Accessed: 08 March 2025.

\bibitem{mazeika2024harmbench}
Mantas Mazeika, Long Phan, Xuwang Yin, Andy Zou, Zifan Wang, Norman Mu, Elham Sakhaee, Nathaniel Li, Steven Basart, Bo~Li, et~al.
\newblock Harmbench: A standardized evaluation framework for automated red teaming and robust refusal.
\newblock {\em arXiv preprint arXiv:2402.04249}, 2024.

\bibitem{porter2022distinguishing}
Zoe Porter, Annette Zimmermann, Phillip Morgan, John McDermid, Tom Lawton, and Ibrahim Habli.
\newblock Distinguishing two features of accountability for ai technologies.
\newblock {\em Nature Machine Intelligence}, 4(9):734--736, 2022.

\bibitem{raji2020closing}
Inioluwa~Deborah Raji, Andrew Smart, Rebecca~N White, Margaret Mitchell, Timnit Gebru, Ben Hutchinson, Jamila Smith-Loud, Daniel Theron, and Parker Barnes.
\newblock Closing the ai accountability gap: Defining an end-to-end framework for internal algorithmic auditing.
\newblock In {\em Proceedings of the 2020 conference on fairness, accountability, and transparency}, pages 33--44, 2020.

\bibitem{cobbe2023understanding}
Jennifer Cobbe, Michael Veale, and Jatinder Singh.
\newblock Understanding accountability in algorithmic supply chains.
\newblock In {\em Proceedings of the 2023 ACM Conference on Fairness, Accountability, and Transparency}, pages 1186--1197, 2023.

\bibitem{carlan2024dynamic}
Carmen C{\^a}rlan, Francesca Gomez, Yohan Mathew, Ketana Krishna, Ren{\'e} King, Peter Gebauer, and Ben~R Smith.
\newblock Dynamic safety cases for frontier ai.
\newblock {\em arXiv preprint arXiv:2412.17618}, 2024.

\end{thebibliography}

\end{document}